\documentclass[twocolumn]{aastex63}
\usepackage{graphicx}
\usepackage{ulem}
\usepackage{cellspace}
\usepackage{amssymb}
\usepackage{amsmath}
\usepackage{longtable}
\usepackage{tabu}
\usepackage{color}
\usepackage{etoolbox}
\usepackage{supertabular}
\usepackage{savesym}
\savesymbol{tablenum}
\usepackage{siunitx}
\restoresymbol{SIX}{tablenum}
\usepackage{dcolumn}
\usepackage{ wasysym }

\def\revised#1{{{\color{black}#1}}}

\newcommand\Tstrut{\rule{0pt}{2.9ex}}       
\newcommand\Bstrut{\rule[-1.3ex]{0pt}{0pt}} 
\newcommand\TBstrut{\Tstrut\Bstrut}

\newcommand{\deltaf}{2.6 \times 10^{-9}}
\newcommand{\deltafdot}{1.9 \times 10^{-17} }
\newcommand{\freqShortJzeroonefivefour}{\approx 845.8}

\newcommand{\freqShortJzerofivezeronine}{\approx 493.1}
\newcommand{\freqShortJzerosevenzeronine}{\approx  58.1}
\newcommand{\freqShortJzeroseventhreetwo}{\approx 489.0}
\newcommand{\freqShortJzeroeighttwofour}{\approx 202.8}
\newcommand{\freqShortJonefouroneone}{\approx  32.0}
\newcommand{\freqShortJtwotwozerofour}{\approx  23.6}

\newcommand{\AmplUnrestULJzeroonefivefour}{\num{2.7}}
\newcommand{\AmplUnrestULexpJzeroonefivefour}{\num{-26}}
\newcommand{\AmplUnrestULerrpJzeroonefivefour}{0.5}
\newcommand{\AmplUnrestULerrmJzeroonefivefour}{0.5}

\newcommand{\epsUnrestULJzeroonefivefour}{\num{3.1e-08}}

\newcommand{\AmplRatioUnresJzeroonefivefour}{\num{30.3}}

\newcommand{\AmplSpndwnJzeroonefivefour}{\num{9.0e-28}}
\newcommand{\AmplUnrestULJzerofivezeronine}{\num{1.9}}
\newcommand{\AmplUnrestULexpJzerofivezeronine}{\num{-26}}
\newcommand{\AmplUnrestULerrpJzerofivezeronine}{0.5}
\newcommand{\AmplUnrestULerrmJzerofivezeronine}{0.4}

\newcommand{\epsUnrestULJzerofivezeronine}{\num{1.1e-07}}

\newcommand{\AmplRatioUnresJzerofivezeronine}{\num{36.3}}

\newcommand{\AmplSpndwnJzerofivezeronine}{\num{5.2e-28}}
\newcommand{\AmplUnrestULJzerosevenzeronine}{\num{3.3}}
\newcommand{\AmplUnrestULexpJzerosevenzeronine}{\num{-26}}
\newcommand{\AmplUnrestULerrpJzerosevenzeronine}{0.5}
\newcommand{\AmplUnrestULerrmJzerosevenzeronine}{0.6}
\newcommand{\AmplRestULJzerosevenzeronine}{\num{3.9}}

\newcommand{\AmplRestULerrpJzerosevenzeronine}{0.6}
\newcommand{\AmplRestULerrmJzerosevenzeronine}{0.7}
\newcommand{\epsUnrestULJzerosevenzeronine}{\num{1.6e-05}}
\newcommand{\epsRestULJzerosevenzeronine}{\num{2.0e-05}}
\newcommand{\AmplRatioUnresJzerosevenzeronine}{\num{21.9}}
\newcommand{\AmplRatioResJzerosevenzeronine}{\num{26.5}}

\newcommand{\AmplSpndwnJzerosevenzeronine}{\num{1.5e-27}}
\newcommand{\AmplUnrestULJzeroseventhreetwo}{\num{1.9}}
\newcommand{\AmplUnrestULexpJzeroseventhreetwo}{\num{-26}}
\newcommand{\AmplUnrestULerrpJzeroseventhreetwo}{0.5}
\newcommand{\AmplUnrestULerrmJzeroseventhreetwo}{0.4}
\newcommand{\AmplRestULJzeroseventhreetwo}{\num{1.4}}
\newcommand{\AmplRestULexpJzeroseventhreetwo}{\num{-26}}
\newcommand{\AmplRestULerrpJzeroseventhreetwo}{0.4}
\newcommand{\AmplRestULerrmJzeroseventhreetwo}{0.4}
\newcommand{\epsUnrestULJzeroseventhreetwo}{\num{1.2e-07}}
\newcommand{\epsRestULJzeroseventhreetwo}{\num{9.2e-08}}
\newcommand{\AmplRatioUnresJzeroseventhreetwo}{\num{32.0}}
\newcommand{\AmplRatioResJzeroseventhreetwo}{\num{24.0}}

\newcommand{\AmplSpndwnJzeroseventhreetwo}{\num{5.8e-28}}
\newcommand{\AmplUnrestULJzeroeighttwofour}{\num{1.4}}
\newcommand{\AmplUnrestULexpJzeroeighttwofour}{\num{-26}}
\newcommand{\AmplUnrestULerrpJzeroeighttwofour}{0.3}
\newcommand{\AmplUnrestULerrmJzeroeighttwofour}{0.3}
\newcommand{\AmplRestULJzeroeighttwofour}{\num{1.8}}
\newcommand{\AmplRestULexpJzeroeighttwofour}{\num{-26}}
\newcommand{\AmplRestULerrpJzeroeighttwofour}{0.4}
\newcommand{\AmplRestULerrmJzeroeighttwofour}{0.4}
\newcommand{\epsUnrestULJzeroeighttwofour}{\num{4.9e-07}}
\newcommand{\epsRestULJzeroeighttwofour}{\num{6.2e-07}}
\newcommand{\AmplRatioUnresJzeroeighttwofour}{\num{6.9}}
\newcommand{\AmplRatioResJzeroeighttwofour}{\num{8.8}}

\newcommand{\AmplSpndwnJzeroeighttwofour}{\num{2.0e-27}}
\newcommand{\AmplUnrestULJonefouroneone}{\num{5.2}}
\newcommand{\AmplUnrestULexpJonefouroneone}{\num{-26}}
\newcommand{\AmplUnrestULerrpJonefouroneone}{1.2}
\newcommand{\AmplUnrestULerrmJonefouroneone}{1.4}
\newcommand{\AmplRestULJonefouroneone}{\num{3.6}}

\newcommand{\AmplRestULerrpJonefouroneone}{0.9}
\newcommand{\AmplRestULerrmJonefouroneone}{0.9}
\newcommand{\epsUnrestULJonefouroneone}{\num{4.7e-05}}
\newcommand{\epsRestULJonefouroneone}{\num{3.2e-05}}
\newcommand{\AmplRatioUnresJonefouroneone}{\num{47.7}}
\newcommand{\AmplRatioResJonefouroneone}{\num{32.9}}

\newcommand{\AmplSpndwnJonefouroneone}{\num{1.1e-27}}
\newcommand{\AmplUnrestULJtwotwozerofour}{\num{1.5}}
\newcommand{\AmplUnrestULexpJtwotwozerofour}{\num{-25}}
\newcommand{\AmplUnrestULerrpJtwotwozerofour}{0.4}
\newcommand{\AmplUnrestULerrmJtwotwozerofour}{0.3}

\newcommand{\epsUnrestULJtwotwozerofour}{\num{5.6e-04}}

\newcommand{\AmplRatioUnresJtwotwozerofour}{\num{319.2}}

\newcommand{\AmplSpndwnJtwotwozerofour}{\num{4.8e-28}}

\newcommand{\AmplUnrestOoneOtwoOthreeULJzeroonefivefour}{\num{1.5}}
\newcommand{\AmplUnrestOoneOtwoOthreeULexpJzeroonefivefour}{\num{-26}}
\newcommand{\AmplUnrestOoneOtwoOthreeULerrpJzeroonefivefour}{0.3}
\newcommand{\AmplUnrestOoneOtwoOthreeULerrmJzeroonefivefour}{0.3}

\newcommand{\epsUnrestOoneOtwoOthreeULJzeroonefivefour}{\num{1.7e-08}}

\newcommand{\AmplOoneOtwoOthreeRatioUnresJzeroonefivefour}{\num{16.7}}

\newcommand{\AmplUnrestOoneOtwoOthreeULJzerofivezeronine}{\num{1.0}}
\newcommand{\AmplUnrestOoneOtwoOthreeULexpJzerofivezeronine}{\num{-26}}
\newcommand{\AmplUnrestOoneOtwoOthreeULerrpJzerofivezeronine}{0.2}
\newcommand{\AmplUnrestOoneOtwoOthreeULerrmJzerofivezeronine}{0.3}

\newcommand{\epsUnrestOoneOtwoOthreeULJzerofivezeronine}{\num{5.7e-08}}

\newcommand{\AmplOoneOtwoOthreeRatioUnresJzerofivezeronine}{\num{19.6}}

\newcommand{\AmplUnrestOoneOtwoOthreeULJzerosevenzeronine}{\num{1.5}}
\newcommand{\AmplUnrestOoneOtwoOthreeULexpJzerosevenzeronine}{\num{-26}}
\newcommand{\AmplUnrestOoneOtwoOthreeULerrpJzerosevenzeronine}{0.3}
\newcommand{\AmplUnrestOoneOtwoOthreeULerrmJzerosevenzeronine}{0.3}
\newcommand{\AmplRestOoneOtwoOthreeULJzerosevenzeronine}{\num{1.9}}
\newcommand{\AmplRestOoneOtwoOthreeULexpJzerosevenzeronine}{\num{-26}}
\newcommand{\AmplRestOoneOtwoOthreeULerrpJzerosevenzeronine}{0.3}
\newcommand{\AmplRestOoneOtwoOthreeULerrmJzerosevenzeronine}{0.4}
\newcommand{\epsUnrestOoneOtwoOthreeULJzerosevenzeronine}{\num{7.7e-06}}
\newcommand{\epsRestOoneOtwoOthreeULJzerosevenzeronine}{\num{9.5e-06}}
\newcommand{\AmplOoneOtwoOthreeRatioUnresJzerosevenzeronine}{\num{10.3}}
\newcommand{\AmplOoneOtwoOthreeRatioResJzerosevenzeronine}{\num{12.7}}
\newcommand{\AmplUnrestOoneOtwoOthreeULJzeroseventhreetwo}{\num{1.0}}
\newcommand{\AmplUnrestOoneOtwoOthreeULexpJzeroseventhreetwo}{\num{-26}}
\newcommand{\AmplUnrestOoneOtwoOthreeULerrpJzeroseventhreetwo}{0.3}
\newcommand{\AmplUnrestOoneOtwoOthreeULerrmJzeroseventhreetwo}{0.3}
\newcommand{\AmplRestOoneOtwoOthreeULJzeroseventhreetwo}{\num{7.8}}
\newcommand{\AmplRestOoneOtwoOthreeULexpJzeroseventhreetwo}{\num{-27}}
\newcommand{\AmplRestOoneOtwoOthreeULerrpJzeroseventhreetwo}{2.5}
\newcommand{\AmplRestOoneOtwoOthreeULerrmJzeroseventhreetwo}{2.0}
\newcommand{\epsUnrestOoneOtwoOthreeULJzeroseventhreetwo}{\num{6.7e-08}}
\newcommand{\epsRestOoneOtwoOthreeULJzeroseventhreetwo}{\num{5.1e-08}}
\newcommand{\AmplOoneOtwoOthreeRatioUnresJzeroseventhreetwo}{\num{17.6}}
\newcommand{\AmplOoneOtwoOthreeRatioResJzeroseventhreetwo}{\num{13.3}}
\newcommand{\AmplUnrestOoneOtwoOthreeULJzeroeighttwofour}{\num{7.6}}
\newcommand{\AmplUnrestOoneOtwoOthreeULexpJzeroeighttwofour}{\num{-27}}
\newcommand{\AmplUnrestOoneOtwoOthreeULerrpJzeroeighttwofour}{1.6}
\newcommand{\AmplUnrestOoneOtwoOthreeULerrmJzeroeighttwofour}{1.9}
\newcommand{\AmplRestOoneOtwoOthreeULJzeroeighttwofour}{\num{1.2}}
\newcommand{\AmplRestOoneOtwoOthreeULexpJzeroeighttwofour}{\num{-26}}
\newcommand{\AmplRestOoneOtwoOthreeULerrpJzeroeighttwofour}{0.3}
\newcommand{\AmplRestOoneOtwoOthreeULerrmJzeroeighttwofour}{0.2}
\newcommand{\epsUnrestOoneOtwoOthreeULJzeroeighttwofour}{\num{2.7e-07}}
\newcommand{\epsRestOoneOtwoOthreeULJzeroeighttwofour}{\num{4.1e-07}}
\newcommand{\AmplOoneOtwoOthreeRatioUnresJzeroeighttwofour}{\num{3.8}}
\newcommand{\AmplOoneOtwoOthreeRatioResJzeroeighttwofour}{\num{5.8}}
\newcommand{\AmplUnrestOoneOtwoOthreeULJonefouroneone}{\num{3.1}}
\newcommand{\AmplUnrestOoneOtwoOthreeULexpJonefouroneone}{\num{-26}}
\newcommand{\AmplUnrestOoneOtwoOthreeULerrpJonefouroneone}{0.7}
\newcommand{\AmplUnrestOoneOtwoOthreeULerrmJonefouroneone}{0.7}
\newcommand{\AmplRestOoneOtwoOthreeULJonefouroneone}{\num{2.2}}
\newcommand{\AmplRestOoneOtwoOthreeULexpJonefouroneone}{\num{-26}}
\newcommand{\AmplRestOoneOtwoOthreeULerrpJonefouroneone}{0.4}
\newcommand{\AmplRestOoneOtwoOthreeULerrmJonefouroneone}{0.3}
\newcommand{\epsUnrestOoneOtwoOthreeULJonefouroneone}{\num{2.8e-05}}
\newcommand{\epsRestOoneOtwoOthreeULJonefouroneone}{\num{2.0e-05}}
\newcommand{\AmplOoneOtwoOthreeRatioUnresJonefouroneone}{\num{28.0}}
\newcommand{\AmplOoneOtwoOthreeRatioResJonefouroneone}{\num{19.9}}
\newcommand{\AmplUnrestOoneOtwoOthreeULJtwotwozerofour}{\num{8.1}}
\newcommand{\AmplUnrestOoneOtwoOthreeULexpJtwotwozerofour}{\num{-26}}
\newcommand{\AmplUnrestOoneOtwoOthreeULerrpJtwotwozerofour}{2.0}
\newcommand{\AmplUnrestOoneOtwoOthreeULerrmJtwotwozerofour}{2.1}

\newcommand{\epsUnrestOoneOtwoOthreeULJtwotwozerofour}{\num{2.9e-04}}

\newcommand{\AmplOoneOtwoOthreeRatioUnresJtwotwozerofour}{\num{166.4}}

\newcommand{\AmplUnrestOthreeULJzeroonefivefour}{\num{1.9}}
\newcommand{\AmplUnrestOthreeULexpJzeroonefivefour}{\num{-26}}
\newcommand{\AmplUnrestOthreeULerrpJzeroonefivefour}{0.4}
\newcommand{\AmplUnrestOthreeULerrmJzeroonefivefour}{0.3}

\newcommand{\epsUnrestOthreeULJzeroonefivefour}{\num{2.2e-08}}

\newcommand{\AmplOthreeRatioUnresJzeroonefivefour}{\num{21.2}}

\newcommand{\AmplUnrestOthreeULJzerofivezeronine}{\num{1.2}}
\newcommand{\AmplUnrestOthreeULexpJzerofivezeronine}{\num{-26}}
\newcommand{\AmplUnrestOthreeULerrpJzerofivezeronine}{0.2}
\newcommand{\AmplUnrestOthreeULerrmJzerofivezeronine}{0.3}

\newcommand{\epsUnrestOthreeULJzerofivezeronine}{\num{6.9e-08}}

\newcommand{\AmplOthreeRatioUnresJzerofivezeronine}{\num{23.7}}

\newcommand{\AmplUnrestOthreeULJzerosevenzeronine}{\num{2.2}}
\newcommand{\AmplUnrestOthreeULexpJzerosevenzeronine}{\num{-26}}
\newcommand{\AmplUnrestOthreeULerrpJzerosevenzeronine}{0.3}
\newcommand{\AmplUnrestOthreeULerrmJzerosevenzeronine}{0.4}
\newcommand{\AmplRestOthreeULJzerosevenzeronine}{\num{2.4}}
\newcommand{\AmplRestOthreeULexpJzerosevenzeronine}{\num{-26}}
\newcommand{\AmplRestOthreeULerrpJzerosevenzeronine}{0.3}
\newcommand{\AmplRestOthreeULerrmJzerosevenzeronine}{0.1}
\newcommand{\epsUnrestOthreeULJzerosevenzeronine}{\num{1.1e-05}}
\newcommand{\epsRestOthreeULJzerosevenzeronine}{\num{1.2e-05}}
\newcommand{\AmplOthreeRatioUnresJzerosevenzeronine}{\num{14.6}}
\newcommand{\AmplOthreeRatioResJzerosevenzeronine}{\num{16.1}}
\newcommand{\AmplUnrestOthreeULJzeroseventhreetwo}{\num{1.2}}
\newcommand{\AmplUnrestOthreeULexpJzeroseventhreetwo}{\num{-26}}
\newcommand{\AmplUnrestOthreeULerrpJzeroseventhreetwo}{0.3}
\newcommand{\AmplUnrestOthreeULerrmJzeroseventhreetwo}{0.3}
\newcommand{\AmplRestOthreeULJzeroseventhreetwo}{\num{10.0}}
\newcommand{\AmplRestOthreeULexpJzeroseventhreetwo}{\num{-27}}
\newcommand{\AmplRestOthreeULerrpJzeroseventhreetwo}{2.5}
\newcommand{\AmplRestOthreeULerrmJzeroseventhreetwo}{2.5}
\newcommand{\epsUnrestOthreeULJzeroseventhreetwo}{\num{7.9e-08}}
\newcommand{\epsRestOthreeULJzeroseventhreetwo}{\num{6.6e-08}}
\newcommand{\AmplOthreeRatioUnresJzeroseventhreetwo}{\num{20.6}}
\newcommand{\AmplOthreeRatioResJzeroseventhreetwo}{\num{17.1}}
\newcommand{\AmplUnrestOthreeULJzeroeighttwofour}{\num{9.9}}
\newcommand{\AmplUnrestOthreeULexpJzeroeighttwofour}{\num{-27}}
\newcommand{\AmplUnrestOthreeULerrpJzeroeighttwofour}{1.5}
\newcommand{\AmplUnrestOthreeULerrmJzeroeighttwofour}{1.4}
\newcommand{\AmplRestOthreeULJzeroeighttwofour}{\num{1.2}}
\newcommand{\AmplRestOthreeULexpJzeroeighttwofour}{\num{-26}}
\newcommand{\AmplRestOthreeULerrpJzeroeighttwofour}{0.2}
\newcommand{\AmplRestOthreeULerrmJzeroeighttwofour}{0.3}
\newcommand{\epsUnrestOthreeULJzeroeighttwofour}{\num{3.5e-07}}
\newcommand{\epsRestOthreeULJzeroeighttwofour}{\num{4.2e-07}}
\newcommand{\AmplOthreeRatioUnresJzeroeighttwofour}{\num{5.0}}
\newcommand{\AmplOthreeRatioResJzeroeighttwofour}{\num{5.9}}
\newcommand{\AmplUnrestOthreeULJonefouroneone}{\num{3.9}}
\newcommand{\AmplUnrestOthreeULexpJonefouroneone}{\num{-26}}
\newcommand{\AmplUnrestOthreeULerrpJonefouroneone}{1.1}
\newcommand{\AmplUnrestOthreeULerrmJonefouroneone}{1.1}
\newcommand{\AmplRestOthreeULJonefouroneone}{\num{2.7}}
\newcommand{\AmplRestOthreeULexpJonefouroneone}{\num{-26}}
\newcommand{\AmplRestOthreeULerrpJonefouroneone}{0.7}
\newcommand{\AmplRestOthreeULerrmJonefouroneone}{0.7}
\newcommand{\epsUnrestOthreeULJonefouroneone}{\num{3.5e-05}}
\newcommand{\epsRestOthreeULJonefouroneone}{\num{2.5e-05}}
\newcommand{\AmplOthreeRatioUnresJonefouroneone}{\num{35.7}}
\newcommand{\AmplOthreeRatioResJonefouroneone}{\num{25.1}}
\newcommand{\AmplUnrestOthreeULJtwotwozerofour}{\num{9.9}}
\newcommand{\AmplUnrestOthreeULexpJtwotwozerofour}{\num{-26}}
\newcommand{\AmplUnrestOthreeULerrpJtwotwozerofour}{2.6}
\newcommand{\AmplUnrestOthreeULerrmJtwotwozerofour}{2.4}

\newcommand{\epsUnrestOthreeULJtwotwozerofour}{\num{3.6e-04}}

\newcommand{\AmplOthreeRatioUnresJtwotwozerofour}{\num{204.0}}

\newcommand{\raJzeroonefivefour}{0.5001010}
\newcommand{\decJzeroonefivefour}{0.3240047}

\newcommand{\fdotJzeroonefivefour}{\num{-1.046e-15}}
\newcommand{\pepochJzeroonefivefour}{\num{56900.0}}

\newcommand{\raJzerofivezeronine}{1.3498838}
\newcommand{\decJzerofivezeronine}{0.1560374}

\newcommand{\fdotJzerofivezeronine}{\num{-5.366e-16}}
\newcommand{\pepochJzerofivezeronine}{\num{57384.0}}

\newcommand{\asiniJzerofivezeronine}{\num{2.458025534}}
\newcommand{\eccJzerofivezeronine}{ 0.00002}
\newcommand{\pbJzerofivezeronine}{\num{424049.2036136159}}
\newcommand{\omegaJzerofivezeronine}{0.5642}
\newcommand{\raJzerosevenzeronine}{1.8724740}
\newcommand{\decJzerosevenzeronine}{0.0869343}

\newcommand{\fdotJzerosevenzeronine}{\num{-6.418e-16}}
\newcommand{\pepochJzerosevenzeronine}{\num{56983.893691}}

\newcommand{\asiniJzerosevenzeronine}{\num{15.716582025}}
\newcommand{\eccJzerosevenzeronine}{ 0.00023}
\newcommand{\pbJzerosevenzeronine}{\num{377281.1771422851}}
\newcommand{\omegaJzerosevenzeronine}{5.6319}
\newcommand{\iotaJzerosevenzeronine}{1.30}
\newcommand{\raJzeroseventhreetwo}{1.9749503}
\newcommand{\decJzeroseventhreetwo}{0.4057610}

\newcommand{\fdotJzeroseventhreetwo}{\num{-7.344e-16}}
\newcommand{\pepochJzeroseventhreetwo}{\num{58000.0}}

\newcommand{\asiniJzeroseventhreetwo}{\num{10.625842295}}
\newcommand{\eccJzeroseventhreetwo}{ 0.00001}
\newcommand{\pbJzeroseventhreetwo}{\num{2611878.6842582175}}
\newcommand{\omegaJzeroseventhreetwo}{1.1879}
\newcommand{\iotaJzeroseventhreetwo}{0.93}
\newcommand{\raJzeroeighttwofour}{2.2009213}
\newcommand{\decJzeroeighttwofour}{0.0081476}

\newcommand{\fdotJzeroeighttwofour}{\num{-3.021e-15}}
\newcommand{\pepochJzeroeighttwofour}{\num{56600.0}}

\newcommand{\asiniJzeroeighttwofour}{\num{18.988928488}}
\newcommand{\eccJzeroeighttwofour}{ 0.00023}
\newcommand{\pbJzeroeighttwofour}{\num{2005081.527370442}}
\newcommand{\omegaJzeroeighttwofour}{0.8084}
\newcommand{\iotaJzeroeighttwofour}{1.32}
\newcommand{\raJonefouroneone}{3.7145600}
\newcommand{\decJonefouroneone}{0.4512083}

\newcommand{\fdotJonefouroneone}{\num{-4.904e-17}}
\newcommand{\pepochJonefouroneone}{\num{57257.864168}}

\newcommand{\asiniJonefouroneone}{\num{9.204790917}}
\newcommand{\eccJonefouroneone}{ 0.16993}
\newcommand{\pbJonefouroneone}{\num{226010.02575114663}}
\newcommand{\omegaJonefouroneone}{1.4209}
\newcommand{\iotaJonefouroneone}{0.83}
\newcommand{\raJtwotwozerofour}{5.7802112}
\newcommand{\decJtwotwozerofour}{0.4715040}

\newcommand{\fdotJtwotwozerofour}{\num{-3.652e-17}}
\newcommand{\pepochJtwotwozerofour}{\num{56805.0}}

\newcommand{\asiniJtwotwozerofour}{\num{210.680632593}}
\newcommand{\eccJtwotwozerofour}{ 0.00152}
\newcommand{\pbJtwotwozerofour}{\num{70437206.11210689}}
\newcommand{\omegaJtwotwozerofour}{0.1118}
\newcommand{\DistJtwotwozerofour}{2150}
\newcommand{\DistJzerofivezeronine}{1450}
\newcommand{\DistJzerosevenzeronine}{1790}
\newcommand{\DistJzeroseventhreetwo}{1660}
\newcommand{\DistJzeroeighttwofour}{1530}
\newcommand{\DistJzeroonefivefour}{860}

\newcommand{\DistJonefouroneone}{977}

\newcommand{\Freq}{f}
\newcommand{\fdot}{{\dot{\Freq}}}

\newcommand{\F}{\mathcal{F}}		

\newcommand{\AmplUnrestOonetwothreeBayesULJzeroonefivefour}{\num{1.3}}
\newcommand{\AmplUnrestOonetwothreeBayesULexpJzeroonefivefour}{\num{-26}}
\newcommand{\epsUnrestOonetwothreeBayesULJzeroonefivefour}{\num{1.5e-08}}
\newcommand{\AmplOonetwothreeBayesRatioUnresJzeroonefivefour}{\num{14.9}}
\newcommand{\AmplUnrestOonetwothreeBayesULJzerofivezeronine}{\num{1.0}}
\newcommand{\AmplUnrestOonetwothreeBayesULexpJzerofivezeronine}{\num{-26}}
\newcommand{\epsUnrestOonetwothreeBayesULJzerofivezeronine}{\num{5.9e-08}}
\newcommand{\AmplOonetwothreeBayesRatioUnresJzerofivezeronine}{\num{20.1}}
\newcommand{\AmplUnrestOonetwothreeBayesULJzerosevenzeronine}{\num{1.9}}
\newcommand{\AmplUnrestOonetwothreeBayesULexpJzerosevenzeronine}{\num{-26}}
\newcommand{\epsUnrestOonetwothreeBayesULJzerosevenzeronine}{\num{9.4e-06}}
\newcommand{\AmplOonetwothreeBayesRatioUnresJzerosevenzeronine}{\num{12.5}}
\newcommand{\AmplUnrestOonetwothreeBayesULJzeroseventhreetwo}{\num{7.7}}
\newcommand{\AmplUnrestOonetwothreeBayesULexpJzeroseventhreetwo}{\num{-27}}
\newcommand{\epsUnrestOonetwothreeBayesULJzeroseventhreetwo}{\num{5.1e-08}}
\newcommand{\AmplOonetwothreeBayesRatioUnresJzeroseventhreetwo}{\num{13.2}}
\newcommand{\AmplUnrestOonetwothreeBayesULJzeroeighttwofour}{\num{9.0}}
\newcommand{\AmplUnrestOonetwothreeBayesULexpJzeroeighttwofour}{\num{-27}}
\newcommand{\epsUnrestOonetwothreeBayesULJzeroeighttwofour}{\num{3.2e-07}}
\newcommand{\AmplOonetwothreeBayesRatioUnresJzeroeighttwofour}{\num{4.5}}
\newcommand{\AmplUnrestOonetwothreeBayesULJonefouroneone}{\num{2.8}}
\newcommand{\AmplUnrestOonetwothreeBayesULexpJonefouroneone}{\num{-26}}
\newcommand{\epsUnrestOonetwothreeBayesULJonefouroneone}{\num{2.5e-05}}
\newcommand{\AmplOonetwothreeBayesRatioUnresJonefouroneone}{\num{25.9}}
\newcommand{\AmplUnrestOonetwothreeBayesULJtwotwozerofour}{\num{7.3}}
\newcommand{\AmplUnrestOonetwothreeBayesULexpJtwotwozerofour}{\num{-26}}
\newcommand{\epsUnrestOonetwothreeBayesULJtwotwozerofour}{\num{2.7e-04}}
\newcommand{\AmplOonetwothreeBayesRatioUnresJtwotwozerofour}{\num{151.4}}

\shorttitle{New searches for continuous gravitational waves from seven fast pulsars}
\shortauthors{Ashok et al.}

\begin{document}

\title{New searches for continuous gravitational waves from seven fast pulsars}

\correspondingauthor{A. Ashok}
\email{anjana.ashok@aei.mpg.de}

\correspondingauthor{M.A. Papa}
\email{maria.alessandra.papa@aei.mpg.de}

\author[0000-0002-8395-957X]{A. Ashok}
\affiliation{Max Planck Institute for Gravitational Physics (Albert Einstein Institute), Callinstrasse 38, 30167 Hannover, Germany}
\affiliation{Leibniz Universit\"at Hannover, D-30167 Hannover, Germany}

\author{B. Beheshtipour}
\affiliation{Max Planck Institute for Gravitational Physics (Albert Einstein Institute), Callinstrasse 38, 30167 Hannover, Germany}
\affiliation{Leibniz Universit\"at Hannover, D-30167 Hannover, Germany}

\author[0000-0002-1007-5298]{M. A. Papa}
\affiliation{Max Planck Institute for Gravitational Physics (Albert Einstein Institute), Callinstrasse 38, 30167 Hannover, Germany}
\affiliation{University of Wisconsin Milwaukee, 3135 N Maryland Ave, Milwaukee, WI 53211, USA}
\affiliation{Leibniz Universit\"at Hannover, D-30167 Hannover, Germany}

\author{P. C. C. Freire}
\affiliation{ Max-Planck-Institut für Radioastronomie, Auf dem Hügel 69, D-53121
Bonn, Germany}

\author{B. Steltner}
\affiliation{Max Planck Institute for Gravitational Physics (Albert Einstein Institute), Callinstrasse 38, 30167 Hannover, Germany}
\affiliation{Leibniz Universit\"at Hannover, D-30167 Hannover, Germany}

\author{B. Machenschalk}
\affiliation{Max Planck Institute for Gravitational Physics (Albert Einstein Institute), Callinstrasse 38, 30167 Hannover, Germany}
\affiliation{Leibniz Universit\"at Hannover, D-30167 Hannover, Germany}

\author{O. Behnke}
\affiliation{Max Planck Institute for Gravitational Physics (Albert Einstein Institute), Callinstrasse 38, 30167 Hannover, Germany}
\affiliation{Leibniz Universit\"at Hannover, D-30167 Hannover, Germany}

\author{B. Allen}
\affiliation{Max Planck Institute for Gravitational Physics (Albert Einstein Institute), Callinstrasse 38, 30167 Hannover, Germany}
\affiliation{University of Wisconsin Milwaukee, 3135 N Maryland Ave, Milwaukee, WI 53211, USA}
\affiliation{Leibniz Universit\"at Hannover, D-30167 Hannover, Germany}

\author{R. Prix}
\affiliation{Max Planck Institute for Gravitational Physics (Albert Einstein Institute), Callinstrasse 38, 30167 Hannover, Germany}
\affiliation{Leibniz Universit\"at Hannover, D-30167 Hannover, Germany}

\begin{abstract}
We conduct searches for continuous gravitational waves from seven pulsars, that have not been targeted in continuous wave searches of Advanced LIGO data before. We target emission at exactly twice the rotation frequency of the pulsars and in a small band around such frequency. The former search assumes that the gravitational wave quadrupole is changing phase-locked with the rotation of the pulsar. The search over a range of frequencies allows for differential rotation between the component emitting the radio signal and the component emitting the gravitational waves, for example the crust or magnetosphere versus the core. Timing solutions derived from the Arecibo 327-MHz  Drift-Scan Pulsar Survey  (AO327) observations are used. No evidence of a signal is found and upper limits are set on the gravitational wave amplitude. For one of the pulsars we probe gravitational wave intrinsic amplitudes just a factor of 3.8 higher than the spin-down limit, assuming a canonical moment of inertia of $10^{38}$ kg m$^2$. Our tightest ellipticity constraint is $1.5 \times 10^{-8}$, which is a value well within the range of what a neutron star crust could support.
\end{abstract}



\section{Introduction}

Continuous gravitational waves are expected from rotating neutron stars if these objects present a deviation from a perfectly axisymmetric configuration \citep{Lasky:2015uia,Jaranowski:1998qm}. On the whole, the expected signal is simple, consisting of one or two harmonics, at the rotation frequency of the star and at twice this frequency \citep{Jones:2015zma}. 

The sensitivity of the LIGO instruments allows to probe continuous gravitational wave emission from the Galactic population of neutron stars, for deformations of a few parts in a million and smaller, depending on the search, over a broad range of frequencies. Different types of searches are carried out: ``blind" all-sky surveys \citep{LIGOScientific:2021tsm,Steltner:2020hfd,Covas:2020nwy,Dergachev:2020upb,Dergachev:2020fli,Pisarski:2019vxw}, searches directed at neutron star candidates like supernova remnants, low mass X-ray binaries \citep{Zhang:2020rph,LIGOScientific:2021mwx,Papa:2020vfz,Lindblom:2020rug,Jones:2020htx,Ming:2019xse} and targeted searches aimed at known pulsars \citep{Authors:2019ztc,Abbott:2019bed,Nieder:2020yqy,Nieder:2019cyc,Fesik:2020tvn,LIGOScientific:2020lkw,LIGOScientific:2021yby}.

Among the different searches, the ones that target pulsars, have a special place. Pulsars are believed to be neutron stars, the distance is usually known and the rotation frequency and its derivatives are also known. This has important consequences: a null measurement is directly informative on the gravitational wave emission -- there is no question about whether a source is there in the first place. The search is simple because whatever the emission mechanism is, the gravitational frequency depends on the spin frequency, which is known. A detection would therefore immediately encode information on what is sourcing the gravitational waves. Because  there is little to no uncertainty on the gravitational waveform from a known pulsar, the number of templates that are searched is many orders of magnitude smaller than those investigated in surveys: \revised{{the O2 data all-sky search of  \cite{Steltner:2020hfd} probed $\approx 10^{17}$ more templates than a targeted search. Fewer probed waveforms make targeted searches the most sensitive: the smallest detectable signal is a few times smaller than what the most sensitive broad survey could detect at the same frequency. }}

In this paper we present results from searches for emission from seven new pulsars using public data from all three Advanced LIGO observing runs O1, O2 and O3 \citep{Abbott:2021boh,o1_data,o2_data,o3a_data}.

The plan of the paper is the following: we introduce the signal model in Section \ref{sec:signal}. In Sections \ref{sec:pulsars} we detail the targeted objects. The gravitational wave searches are described in \ref{sec:SearchGeneral}, the results are presented and discussed in \ref{sec:results}. 

\section{The signal}
\label{sec:signal}

The search described in this paper targets nearly monochromatic gravitational wave signals of the form described for example in Section II of  \cite{Jaranowski:1998qm}.  In the calibrated strain data from a gravitational wave detector the signal has the form
\begin{equation}
h(t)=F_+ (\alpha,\delta,\psi ;t) h_+ (t) + F_\times (\alpha,\delta,\psi; t) h_\times(t), 
\label{eq:signal}
\end{equation}
with the ``+" and ``$\times$" indicating the two gravitational wave polarizations.  $F_+ (\alpha,\delta,\psi;t)$ and
$F_\times(\alpha,\delta,\psi;t)$ are the detector sensitivity pattern functions, which depend on relative orientation between the detector and the source, and hence on time $t$, on the position $(\alpha,\delta)$ of the source, and on $\psi$, the polarization angle.
The waveforms $h_+ (t)$ and $h_\times (t)$ are
\begin{eqnarray}
h_+ (t)  =  A_+ \cos \Phi(t) \nonumber \\
h_\times (t)  =  A_\times \sin \Phi(t),
\label{eq:monochromatic}
\end{eqnarray}
with 
\begin{eqnarray}
A_+  & = & {1\over 2} h_0 (1+\cos^2\iota) \nonumber \\
A_\times & = &  h_0  \cos\iota. 
\label{eq:amplitudes}
\end{eqnarray}
The angle between the total angular momentum of the star and the line of sight is $0\leq \iota \leq \pi$ and 
$h_0\geq 0$ is the intrinsic gravitational wave amplitude. 
$\Phi(t)$ of Eq.~\ref{eq:monochromatic} is the phase of the gravitational wave signal at time
$t$. If $\tau_{\mathrm{SSB}}$ is the arrival time of the wave with phase $\Phi(t)$ at the solar system barycenter, then $\Phi(t)=\Phi(\tau_{\mathrm{SSB}}(t))$. The gravitational wave phase as function of $\tau_{\mathrm{SSB}}$ is assumed to be 
\begin{multline}
\label{eq:phiSSB}
\Phi(\tau_{\mathrm{SSB}}) = \Phi_0 + 2\pi [ f(\tau_{\mathrm{SSB}}-{\tau_0}_{\mathrm{SSB}})  +
\\ {1\over 2} \dot{f} (\tau_{\mathrm{SSB}}-{\tau_0}_{\mathrm{SSB}})^2 ].
\end{multline}
We take ${\tau_0}_{\mathrm{SSB}}$ consistently with the timing solution, and hence different for every pulsar, as shown in Table \ref{tab:pulsarParams}. 

\section{The Pulsars}
\label{sec:pulsars}

We target continuous gravitational wave emission from seven recycled pulsars discovered and/or timed with data from the Arecibo 327-MHz  Drift-Scan Pulsar Survey  (AO327) \citep{Martinez:2019thh,Martinez:2017jbp}: PSR J2204+2700, PSR J1411+2551, PSR J0709+0458, PSR J0824+0028, PSR J0732+2314,  PSR J0509+0856 and PSR J0154+1833. For practicality we mostly use abbreviated forms of the names of the pulsars, omitting the ``PSR" prefix and the part after the ``+". 

These pulsars have never been searched before, for gravitational wave emission.
They represent a relatively nearby sample, with distances smaller than 2 kpc, which is typical of all-sky surveys. This makes them particularly interesting for gravitational wave searches, the only exception being 2204+2700, which is more distant, and also having an extremely low spindown. 

Our targets are all in binary systems except for J0154+1833, that is an isolated millisecond pulsar. Our set includes the radio pulsar in the notable double-neutron star system PSR J1411+2551. 

When available, we take the orbital inclination angle as estimate of the inclination angle $\iota$ for the determination of the constrained prior upper limit, see Section \ref{sec:ULs}. We take the following values: $\iota_{J1411} = \iotaJonefouroneone$ rad,  $\iota_{J0709} = \iotaJzerosevenzeronine$ rad, $\iota_{J0824} =  \iotaJzeroeighttwofour$ rad, $\iota_{0732} =   \iotaJzeroseventhreetwo$ rad. 
For J0154, J0509 and J2204 we do not have an estimate of the inclination angle.

\section{The Gravitational Wave Searches}
\label{sec:SearchGeneral}

\begin{figure}
\includegraphics[width=0.45\textwidth]{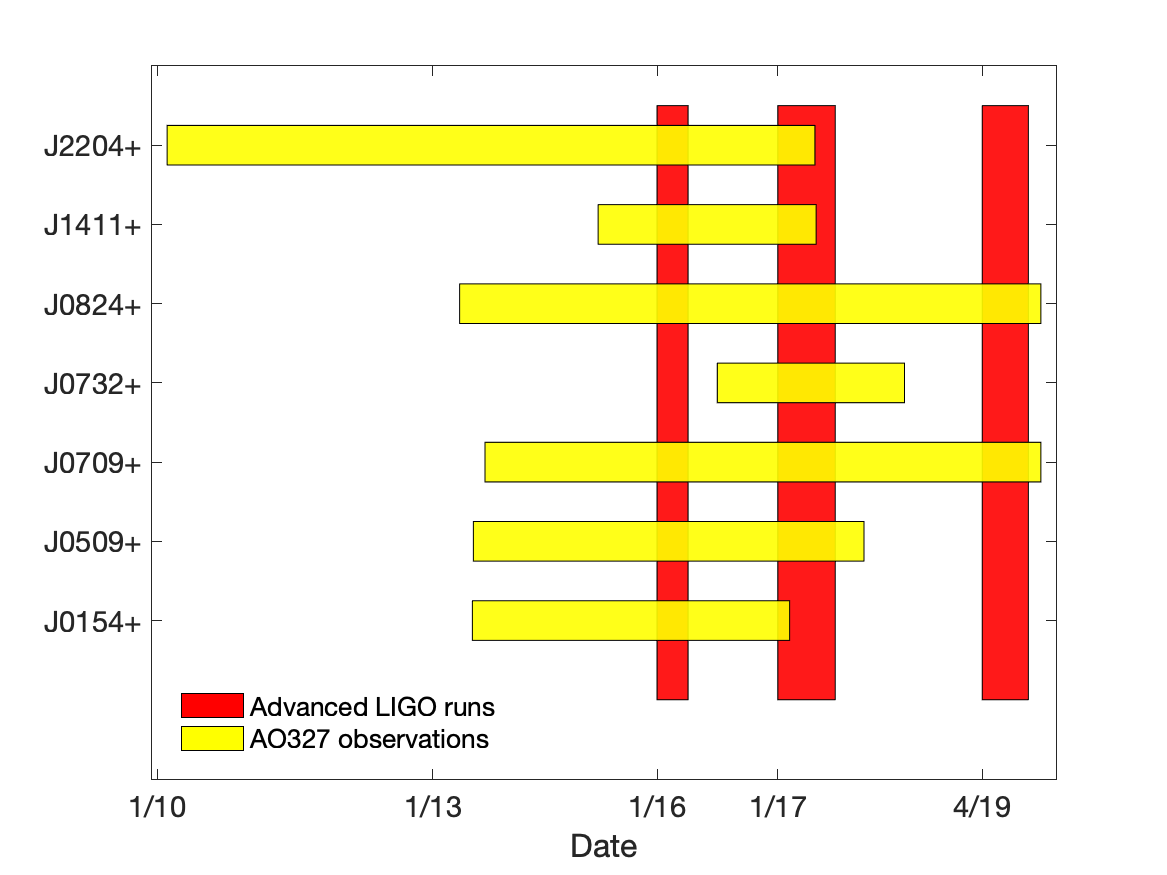}
\caption{Time intervals corresponding to the O1, O2, and O3 LIGO runs are shown in red as ``vertical rectangles" and the radio observation periods for each pulsar are shown in yellow as ``horizontal rectangles".}
\label{fig:emgwoverlap}
\end{figure}

We use LIGO public data from the Hanford (H1) and the Livingston (L1) detectors from the O1, O2 and the recently released first six months of the O3 science run \citep{o1_data,o2_data,o3a_data}. The data is gated to remove loud glitches \citep{Steltner:2021qjy} and contiguous segments are Fourier-transformed to produce the input to the search. After having excluded egregiously noisy segments in the band of each pulsar, we have $\approx$ 175 days of data from each detector from the O1 and O2 runs combined, and $\approx$ 125 during the O3 run for H1  and $\approx$ 129 days for L1. 

No glitch was recorded by the AO327 in any of the pulsars' spins. As Figure~\ref{fig:emgwoverlap} shows, these observations do not perfectly cover all the Advanced LIGO runs so we cannot exclude the possibility of a glitch. Even though our targets are very stable pulsars and a glitch is unlikely, we perform different searches and combine coherently the O1 and O2 data, the O3 data, and also all the data that we have, O1O2O3. We use the matched-filter detection statistic -- $\F$-statistic \citep{Cutler:2005hc} -- as our detection statistic. The $\F$-statistic is the maximum log-likelihood ratio of the signal hypothesis to the Gaussian noise hypothesis. The signal is described by a frequency, spindown, sky-position and orbital parameter values, which define the template waveform and are explicitly searched over. The signal amplitude parameters $\cos\iota$, $\psi$, $\Phi_0$ and $h_0$ are analytically maximised over.

In Gaussian noise the $2\F$-statistic follows a non-central chi-squared distribution with $4$ degrees of freedom, $\chi^2_4(2\F,\rho^2)$. The non-centrality parameter $\rho^2$ is the expected squared signal-to-noise ratio and it is proportional to $h_0^2 T_{\rm{data}}/S_h$, where $T_{\rm{data}}$ is the duration of time for which data is available and $S_h$ is the strain power spectral density of the noise \citep{Jaranowski:1998qm}. 

For every pulsar and data set we conduct two searches: one with a single template with the gravitational wave frequency $f$ and spindown $\fdot$ being  twice the spin frequency $\nu$ and spindown $\dot\nu$, and one for a range of frequencies and spindowns around these. The parameters of the targeted searches are given in Table \ref{tab:pulsarParams} in the appendix.

The search at $f=2\nu$ is appropriate if the gravitational wave frequency is exactly locked with the observed spin frequency. Mechanisms however exist that could produce a small difference between the gravitational wave frequency and twice the spin frequency: a misalignment of the rotation axis with the symmetry axis of the star, causing free precession; or the component responsible for the gravitational wave emission -- for example a solid core -- not spinning as the radio-emitting component. \revised{{For such cases, it has been found that}} $f=2\nu(1\pm\delta_f)$ with $\delta_f \lesssim 10^{-4}$ \citep{Jones:2001yg,Abbott:2008fx}. With this in mind, we conservatively perform searches over a band $\pm 2\nu  \times 2 \cdot 10^{-3}$ of $f=2\nu$, and consistently for $\fdot$.

For the band searches we set up a template grid in frequency and spin-down with spacings of $\deltaf$\, Hz and $\deltafdot$\, Hz/s, respectively. These grids yield a maximum mismatch smaller than 1\% for the O1O2 and O3 searches, and smaller than 8\% for the O1O2O3 searches. 

We also conduct the single-template searches using a Bayesian approach. We demodulate the data according to the expected signal, we heterodyne/downsample the data and then search over the waveform amplitude parameters with a nested sampling algorithm. \revised{The method is exactly the same as used by \cite{LIGOScientific:2019xqs}, with the same uniform angular priors, namely $\Phi_{0} \in [0,\pi]$, $\psi \in [0,\frac{\pi}{2}]$, $\cos \iota \in [-1, 1]$. For the intrinsic amplitude we adopt the same broad uniform prior for all sources with $h_{0} \in [10^{-27},10^{-24}]$.} The data used for this search is not gated. We report the results for the combined O1O2O3 data.

\begin{figure*}
\centering
\includegraphics[width=0.42\textwidth]{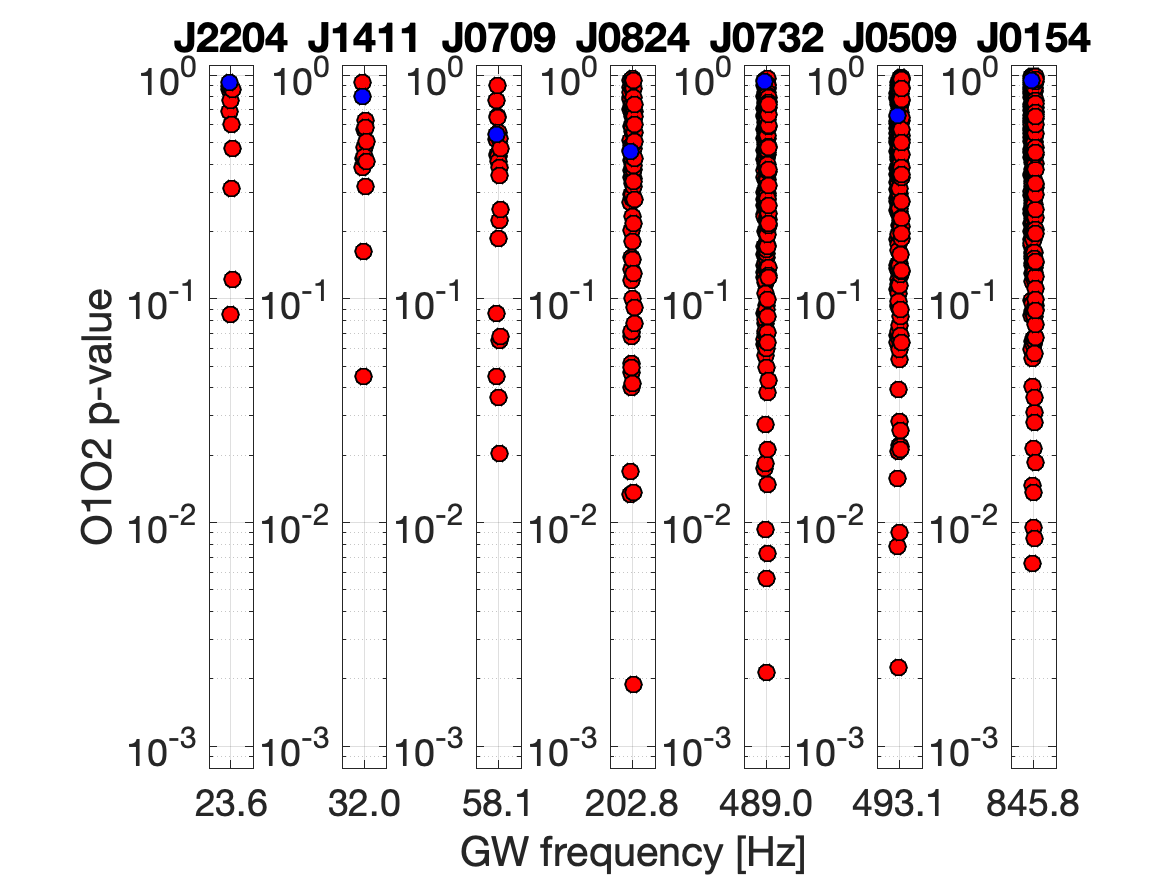}
\includegraphics[width=0.42\textwidth]{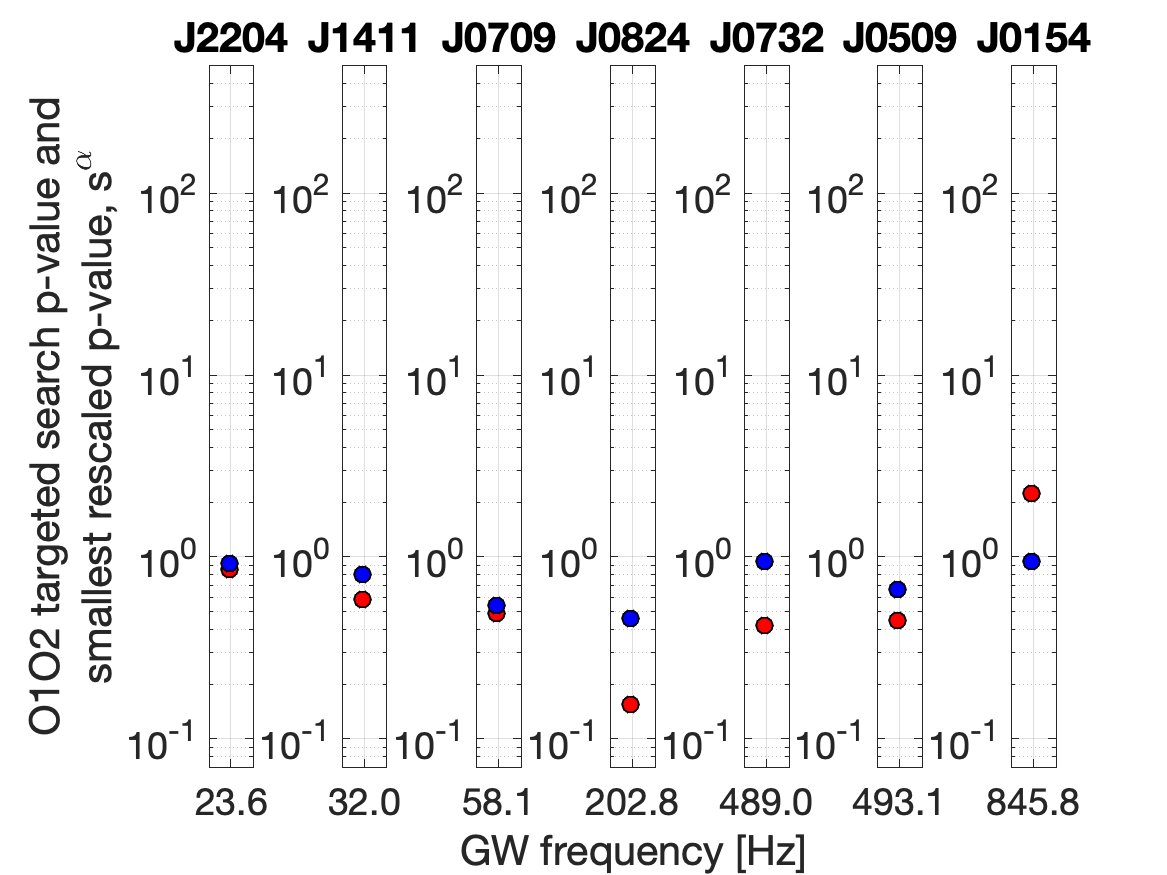}
\includegraphics[width=0.42\textwidth]{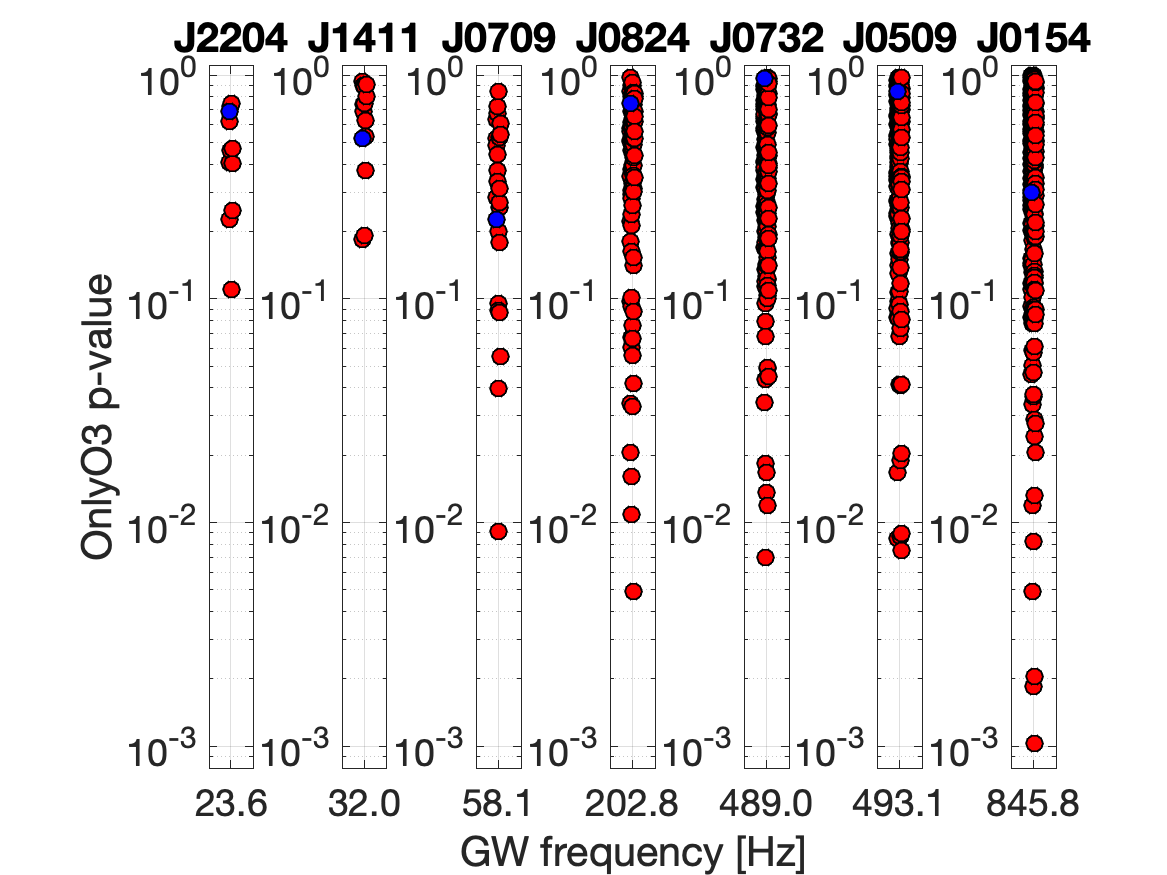}
\includegraphics[width=0.42\textwidth]{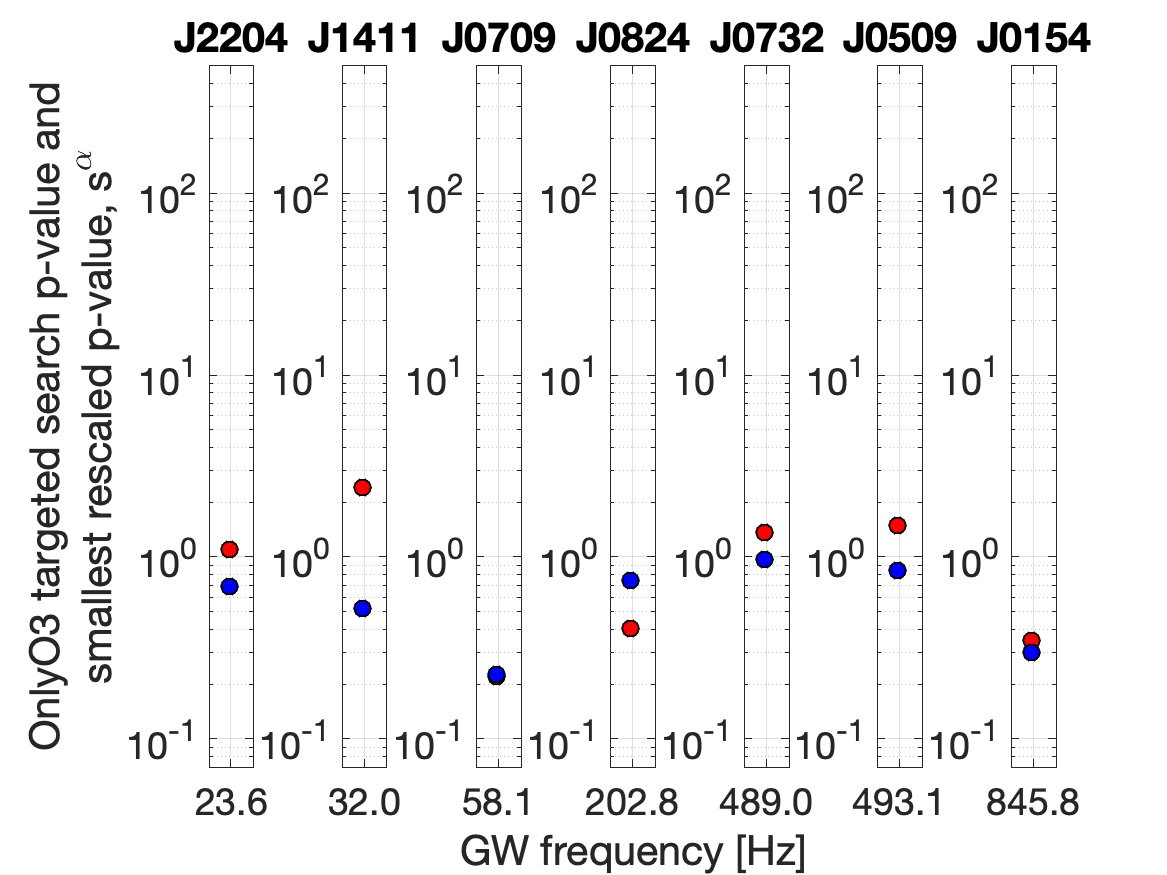}
\includegraphics[width=0.42\textwidth]{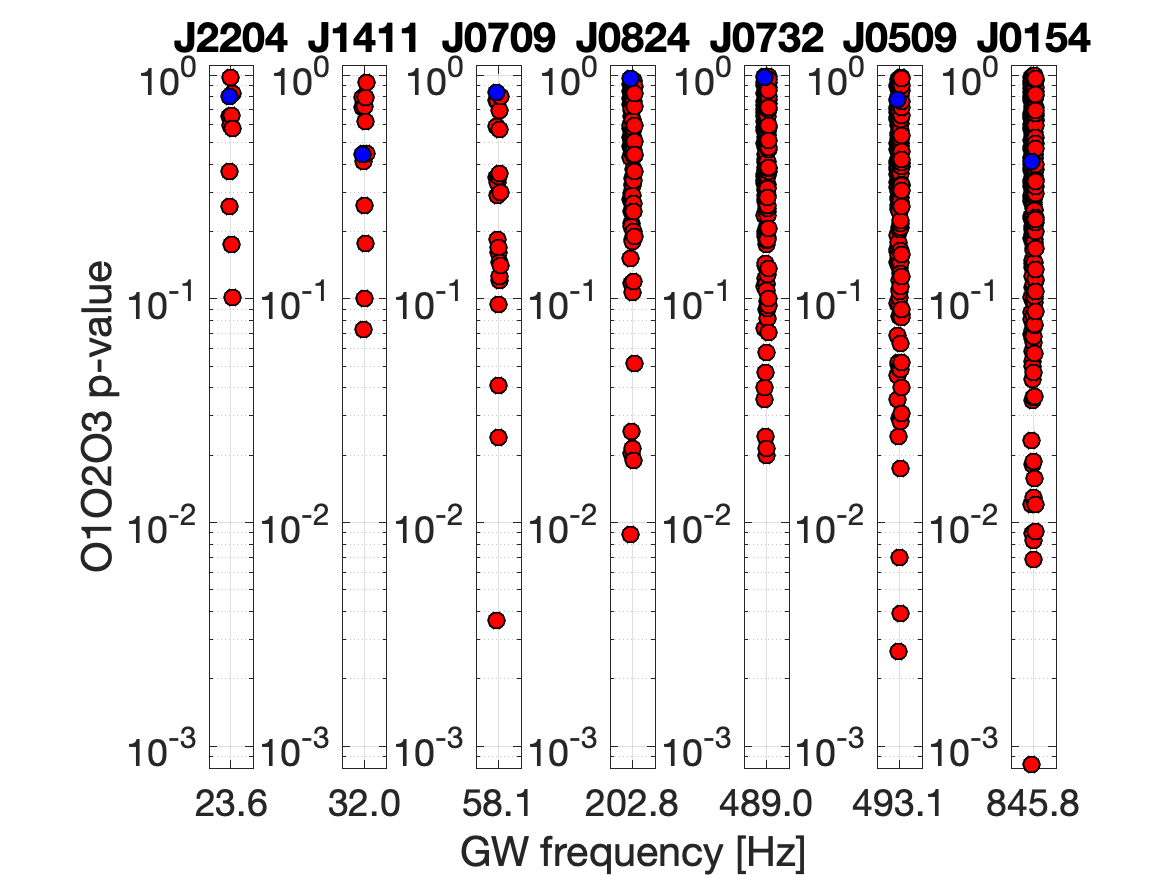}
\includegraphics[width=0.42\textwidth]{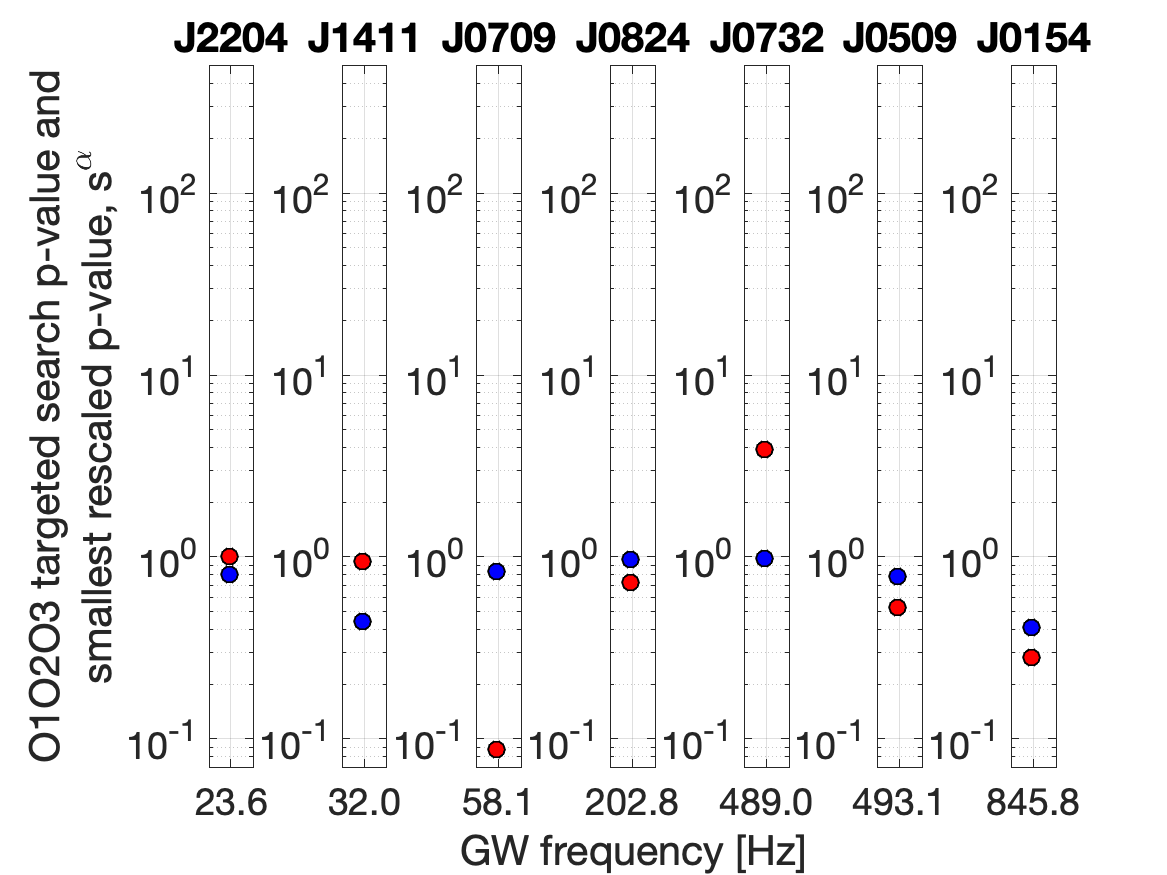}
\caption{O1O2 (1st row), O3 (2nd row) and O1O2O3 (3rd row) results. The blue circles show the p-values of the targeted searches. The red circles in the left-side plots show the p-value of the most significant result in each 10 mHz sub-band of the band searches. These are generally higher than the targeted search p-values because they are maxima over 10 mHz, whereas the targeted searches probe only a single waveform. The red circles in the right-side plots show for each pulsar the lowest p-value amongst the sub-bands rescaled according to Eq.~\ref{eq:rescaledpval}, $s^\alpha$. \revised{{When $s^\alpha \leq 1$ it can be directly interpreted as p-value. When $s^\alpha > 1$ it represents the number of 10 mHz sub-bands in which we'd expect,  in a band search of Gaussian noise data like that performed for pulsar $\alpha$, to measure a result more significant than the most significant found in real data. 
}} }

\label{fig:BandSearchPvals}
\end{figure*}

\begin{figure}
\centering
\includegraphics[width=0.52\textwidth]{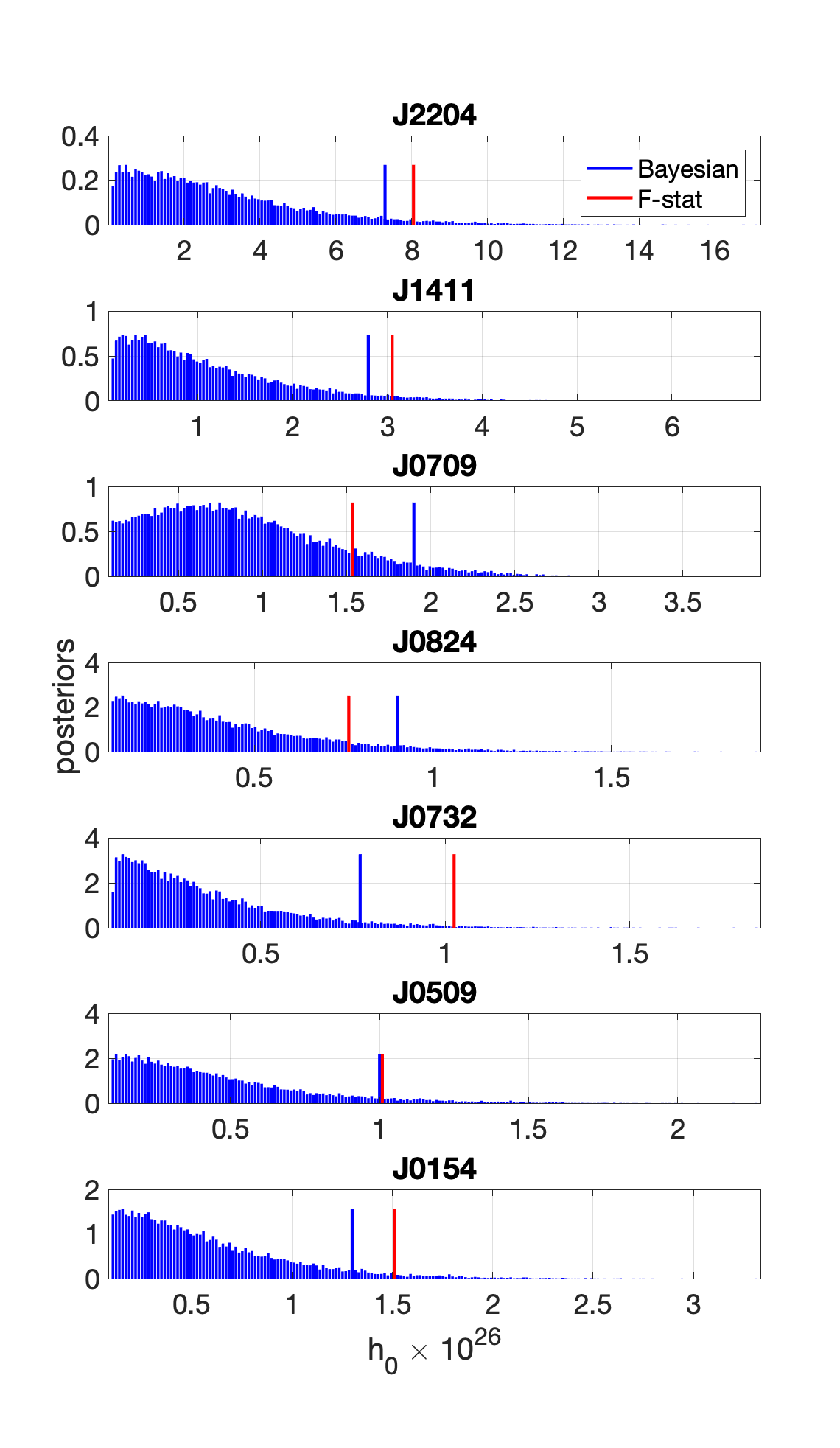}
\caption{Bayesian posteriors for the combined O1O2O3 searches (blue) and associated 95\% confidence upper limits. We also show (red) the $\F$-stat upper limits.}
\label{fig:BayesianULs}
\end{figure}

\begin{figure}
\includegraphics[width=0.45\textwidth]{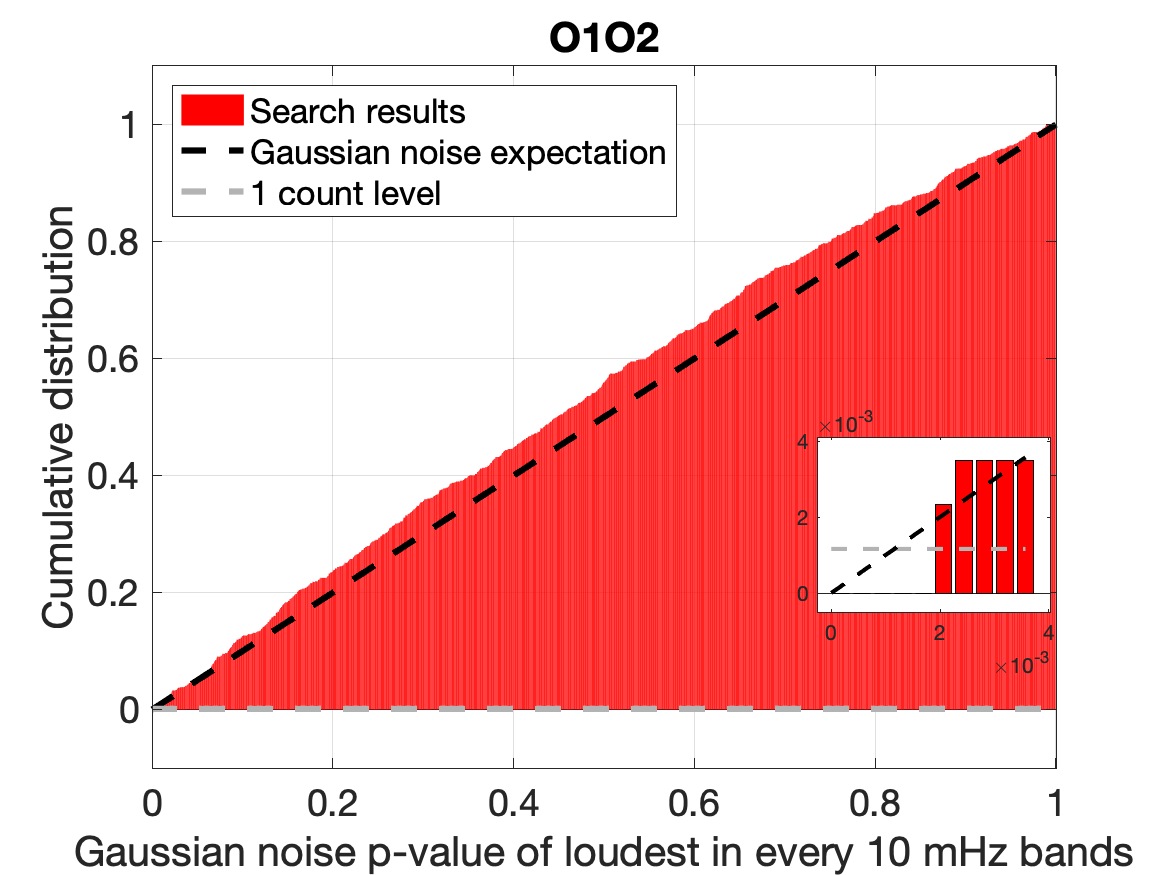}
\includegraphics[width=0.45\textwidth]{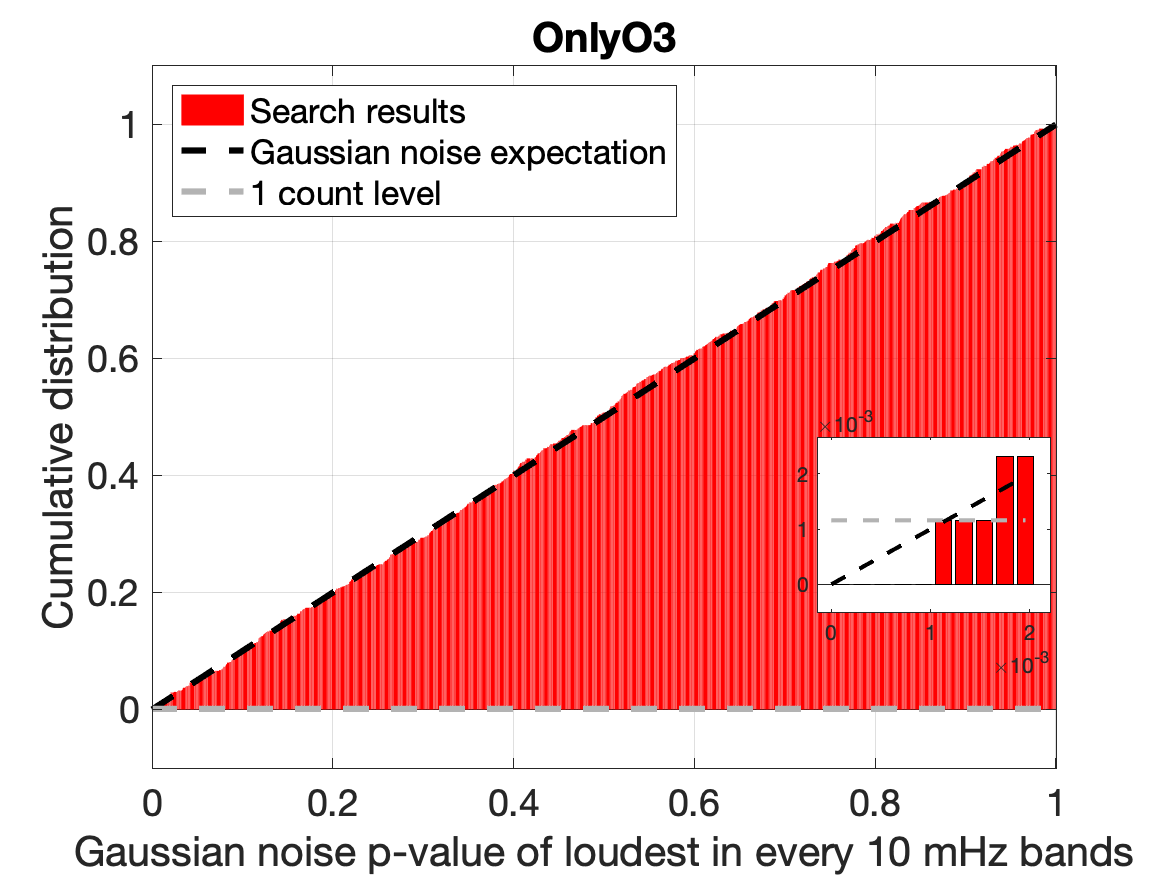}
\includegraphics[width=0.45\textwidth]{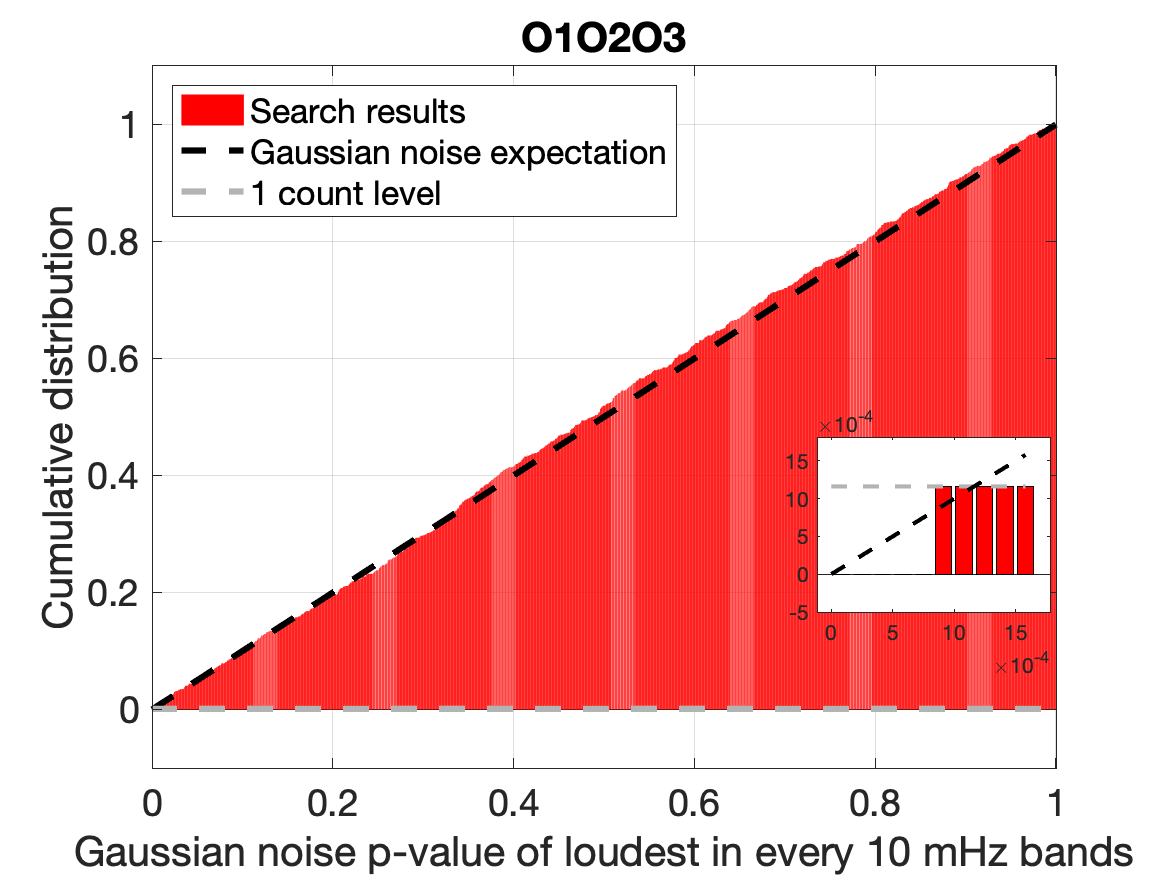}
\caption{O1O2, O3 and O1O2O3 band-search results. For each 10 mHz frequency band searched, we show the cumulative distribution of the Gaussian p-value of the most significant result. If the data were Gaussian noise, the distribution would follow the dashed black line.}
\label{fig:BandSearchPvalHist}
\end{figure}

\section{Results}
\label{sec:results}

\revised{In order to evaluate the significance of the search results we compute p-values. We do this, because based on insignificant p-values we can exclude the presence of signals that we can confidently detect. We note however that a very low p-value in general is not enough to claim a confident detection.}

\revised{The p-value associated with the realisation $2\F^\prime$ of a random variable is defined as }
\begin{equation}
\textrm{p}(2\F^\prime)=\int_{2\F^\prime}^\infty p_0(2\F) ~d 2\F,
\label{eq:pvaldef}
\end{equation}
\revised{where $p_0$ is the distribution of $2\F$ in the presence of noise only.}

\revised{If our data were Gaussian and our search pipelines were completely perfect implementations of the $\F$-statistic, $p_0=\chi^2_4(2\F,0)$. In reality the distribution of our search results may differ slightly from $\chi^2_4(2\F,0)$ and for targeted searches we evaluate it on the actual data, by running searches for fiducial sources at frequencies close to the target frequency.}

None of the targeted searches yield a detection. Figure \ref{fig:BandSearchPvals} shows the p-values for the targeted searches (blue circles): All the results for the targeted searches are consistent with the noise-only hypothesis. The most significant targeted-search result comes from PSR J0709 from the O3-data search, with a p-value of $\approx$ 23\%. The product of the $\approx 55\%$ p-value of the O1O2 result and the O3 result is $\approx 12\%$, however the coherent O1O2O3 data search yields a totally insignificant p-value of $\approx$ 83\% . The Bayesian posteriors of Figure \ref{fig:BayesianULs} are consistent with the $\F$-stat results, with the only slightly off-zero posterior found for PSR J0709. Such posterior is very broad, includes zero and may happen just due to noise fluctuations. We also note that the target frequency for PSR J0709 is at $\approx 58$ Hz which is a highly contaminated region. 

To evaluate the results from the band-searches, for every pulsar we consider the most significant result in sub-bands which are 10 mHz wide, and compute the Gaussian-noise p-value associated with it. \revised{We do this by searching 10 mHz thousands of times, each time with a different Gaussian noise realisation. From each search we find the value of the loudest, 2$\F_\ell$ and from the ensemble we estimate the $p_0(2\F_\ell)$, which we use to compute the p-value from Eq.~\ref{eq:pvaldef}. Since we use Gaussian noise, this is a Gaussian noise p-value and it is conservative (in the sense that we would not accidentally discard a potential signal) because in general it will overestimate the significance of a result with respect to the p-value evaluated on real data. We do not use real data because each Monte Carlo realisation covers 10 mHz; with thousands of independent realisations we'd be considering target frequencies several Hz away from the original pulsar frequency, and at these distances there is no assurance that the noise is representative of the noise contributing to the original result.} These p-values are the red circles shown in the left-hand-side plots in the first three rows of Figure \ref{fig:BandSearchPvals}. 

\revised{{We want to evaluate whether for any pulsar the band-search yields a very significant result, so we consider the most significant (lowest) p-value found for each pulsar. One would want then to compare these lowest p-values but this cannot be done directly, because they do not come from the same distribution. In fact simply due to the trials factor,}} the larger is the band that has been searched, the lower is the expected lowest p-value for that pulsar. \revised{{To normalize the results and allow a direct comparison we estimate the trials factor to be equal to the number of independent 10 mHz sub-bands $N^\alpha$ searched for pulsar $\alpha$, and introduce the following measure of significance for the most significant result for pulsar $\alpha$ :}}
\begin{equation}
\label{eq:rescaledpval}
s^\alpha=N^\alpha~ \underset {i \in [1,N^\alpha] } {\mathrm{min}} \{\textrm{p}^\alpha_i \},
\end{equation}
\revised{{where $\textrm{p}^\alpha_i$ is the p-value associated to the i-th sub-band: $\textrm{p}(2{\F^\star}^\alpha_i)$.}} These quantities are shown in the right-hand-side plots of Figure \ref{fig:BandSearchPvals}. 

\revised{{A value of $s^\alpha < 1$ indicates that we'd have to repeat the $\alpha$-pulsar band search in random noise $1/ s^\alpha$ times before we can expect a result as significant as the observed one.
In this case $s^\alpha$ can be interpreted as a p-value. If $s^\alpha > 1$, it means that in a band search like the one conducted, in random noise, we expect that in $s^\alpha$ sub-bands the loudest values will be at least as significant as the observed result. For a detection we'd need $s^\alpha \lll 1$.}}

The most significant result from the band searches comes again from PSR J0709 in the O1O2O3 coherent search and is at the level of $\approx$ 9\%. However this significance is not confirmed in the O3 data where the lowest p-value ($s^{J0709}$) is 22\% and it is at a different sub-band than the one that produced the O1O2O3 most significant result. 

Figures \ref{fig:BandSearchPvalHist} show the distributions of the most significant 10 mHz p-values and illustrate that they are consistent with the Gaussian-noise expectations for searches on all data. 

\subsection{Upper limits}
\label{sec:ULs}

Based on the O1O2, O3 and O1O2O3 targeted search results we place 95\% confidence upper limits on the intrinsic gravitational wave amplitude at the detector $h_0$ defined in Eq.~\ref{eq:amplitudes}. We use a series of Monte Carlos where we simulate signals at fixed amplitude in real data and measure the detection efficiency of our search. The detection criteria is that the obtained value of the detection statistic be equal or greater than the one found in the real search: if the measured detection statistic is high, a higher gravitational wave amplitude will be needed in order for the signals to be detected. The amplitude for which 95\% of the tested signals is detected, is the upper limit value, $h_0^{95\%}$. With minor variations on the theme, this is the standard approach that we have taken for $\F$-statistic searches since the very first continuous waves search on LIGO data back in \citeyear{Abbott:2003yq}. The $\F$-stat upper limits are shown in Tables~\ref{tab:ULs}. 

\revised{{As discussed in the previous section the $h_0$ posteriors from the Bayesian analysis shown in Fig.~\ref{fig:BayesianULs} are consistent with a null result. Because of this, the Bayesian upper limits are readily derived by integrating the posteriors up to a value such that the overall probability is the desired confidence level. Such value is the upper limit, and it represents the smallest extremum of the credible interval.  The Bayesian upper limit values are shown in Table \ref{tab:BayesianULs}.}}

For the O1O2 band-searches we divide the searched frequency range in 10 mHz sub-bands and take the most significant detection statistic value in that sub-band for our detection criteria. The sub-band searches probe numerous waveforms and so the loudest detection statistic value is going to be higher than for the targeted searches. Correspondingly the upper limits will also be higher as shown in Figure \ref{fig:h0ULs} in the appendix, typically by a factor of $\approx 2.7$. Since this is the most computationally intense part of this work, and we do not find evidence for a signal, we do not set upper limits based on the O3 data, or on the O1O2O3, but we expect that these would be higher than the corresponding targeted ones by also a factor of a few.

The  populations of fake signals used to determine the detection efficiencies have polarization angle $\psi$ and initial phase $\Phi_0$ uniformly distributed as described in Section~\ref{sec:SearchGeneral}. For the orientation angle we consider two cases: $\cos \iota$  uniformly distributed in $[-1,1]$ and fixed at the value of the orbital inclination, when available from the radio observations. We refer to the resulting upper limits as {\it{unconstrained}} and {\it{constrained}}, respectively.

\begin{deluxetable*}{lcccccccc}
\centerwidetable
\tablecaption{95\% confidence upper limits on the gravitational wave amplitude for the targeted searches based on no assumptions on the inclination angle (unconstrained prior) and if the inclination angle is the same as the estimated value of the orbital inclination angle (constrained prior), using different data. We also show the spindown upper limit calculated for a nominal value of the moment of inertia of $10^{38}$ kg m$^2$, and the spindown limit gravitational wave amplitude. The last two columns indicate how far our results are from being physically interesting: if $h_0^{95\%} / h_0^{\textrm{spdwn}}$ is less than one, then the upper limits are informative.
\label{tab:ULs}
}
\tablehead{
\colhead{{Pulsar}} & 
\colhead{$f$ [Hz]} & 
\colhead{$h_0^{95\%}$} & \colhead{$h_0^{95\%}$} & 
\colhead{ $h_0^{\textrm{spdwn}}$ } & 
\colhead{$\epsilon^{95\%}$} & \colhead{$\epsilon^{95\%}$} & 
\colhead{${h_0^{95}}/ {h_0^{\textrm{spdwn}}}$} & \colhead{${h_0^{95}}/ {h_0^{\textrm{spdwn}}}$}\\ [-0.2cm]
\colhead{} & 
\colhead{} &  
\colhead{\tiny{UNCONSTRAINED}} & \colhead{\tiny{CONSTRAINED} } & 
\colhead{} & 
\colhead{\tiny{UNCONSTRAINED}} & \colhead{\tiny{CONSTRAINED}} & 
\colhead{\tiny{UNCONSTRAINED}} & \colhead{\tiny{CONSTRAINED}} \\[-0.3cm]
\colhead{} & 
\colhead{} &  
\colhead{\tiny{PRIOR}} & \colhead{ \tiny{PRIOR}} & 
\colhead{} & 
\colhead{\tiny{PRIOR}} & \colhead{\tiny{PRIOR}} & 
\colhead{\tiny{PRIOR}} & \colhead{\tiny{PRIOR}}
}
\startdata
\cutinhead{O1 O2}
\TBstrut J0154 & $\freqShortJzeroonefivefour$	& 
$\AmplUnrestULJzeroonefivefour^{+\AmplUnrestULerrpJzeroonefivefour}_{-\AmplUnrestULerrmJzeroonefivefour}\times{10^{\AmplUnrestULexpJzeroonefivefour}}$ & 
- & 
$\AmplSpndwnJzeroonefivefour$ 	& 
$\epsUnrestULJzeroonefivefour$ 	& - & 
$\AmplRatioUnresJzeroonefivefour$ & -  \\
\TBstrut J0509 & $\freqShortJzerofivezeronine$ & 
$\AmplUnrestULJzerofivezeronine^{+\AmplUnrestULerrpJzerofivezeronine}_{-\AmplUnrestULerrmJzerofivezeronine}\times{10^{\AmplUnrestULexpJzerofivezeronine}}$ & 
-  & 
$\AmplSpndwnJzerofivezeronine$ 	& 
$\epsUnrestULJzerofivezeronine$ 	& - & 
$\AmplRatioUnresJzerofivezeronine$ & -\\
\TBstrut J0709& $\freqShortJzerosevenzeronine$ & 
$\AmplUnrestULJzerosevenzeronine^{+\AmplUnrestULerrpJzerosevenzeronine}_{-\AmplUnrestULerrmJzerosevenzeronine}\times{10^{\AmplUnrestULexpJzerosevenzeronine}}$ & 
$\AmplRestULJzerosevenzeronine^{+\AmplRestULerrpJzerosevenzeronine}_{-\AmplRestULerrmJzerosevenzeronine}\times{10^{\AmplRestOthreeULexpJzerosevenzeronine}}$  & 
$\AmplSpndwnJzerosevenzeronine$ 	& 
$\epsUnrestULJzerosevenzeronine$ 	& $\epsRestULJzerosevenzeronine$ & 
$\AmplRatioUnresJzerosevenzeronine$ & $\AmplRatioResJzerosevenzeronine$\\
\TBstrut J0732 & $\freqShortJzeroseventhreetwo$ & 
$\AmplUnrestULJzeroseventhreetwo^{+\AmplUnrestULerrpJzeroseventhreetwo}_{-\AmplUnrestULerrmJzeroseventhreetwo}\times{10^{\AmplUnrestULexpJzeroseventhreetwo}}$ & 
$\AmplRestULJzeroseventhreetwo^{+\AmplRestULerrpJzeroseventhreetwo}_{-\AmplRestULerrmJzeroseventhreetwo}\times{10^{\AmplRestULexpJzeroseventhreetwo}}$  & 
$\AmplSpndwnJzeroseventhreetwo$ 	& 
$\epsUnrestULJzeroseventhreetwo$ 	& $\epsRestULJzeroseventhreetwo$ & 
$\AmplRatioUnresJzeroseventhreetwo$ & $\AmplRatioResJzeroseventhreetwo$\\
\TBstrut J0824 & $\freqShortJzeroeighttwofour$ & 
$\AmplUnrestULJzeroeighttwofour^{+\AmplUnrestULerrpJzeroeighttwofour}_{-\AmplUnrestULerrmJzeroeighttwofour}\times{10^{\AmplUnrestULexpJzeroeighttwofour}}$ & 
$\AmplRestULJzeroeighttwofour^{+\AmplRestULerrpJzeroeighttwofour}_{-\AmplRestULerrmJzeroeighttwofour}\times{10^{\AmplRestULexpJzeroeighttwofour}}$  & 
$\AmplSpndwnJzeroeighttwofour$ 	& 
$\epsUnrestULJzeroeighttwofour$ 	& $\epsRestULJzeroeighttwofour$ & 
$\AmplRatioUnresJzeroeighttwofour$ & $\AmplRatioResJzeroeighttwofour$\\
\TBstrut J1411 & $\freqShortJonefouroneone$ & 
$\AmplUnrestULJonefouroneone^{+\AmplUnrestULerrpJonefouroneone}_{-\AmplUnrestULerrmJonefouroneone}\times{10^{\AmplUnrestULexpJonefouroneone}}$ & 
$\AmplRestULJonefouroneone^{+\AmplRestULerrpJonefouroneone}_{-\AmplRestULerrmJonefouroneone}\times{10^{\AmplRestOthreeULexpJonefouroneone}}$  & 
$\AmplSpndwnJonefouroneone$ 	& 
$\epsUnrestULJonefouroneone$ 	& $\epsRestULJonefouroneone$ & 
$\AmplRatioUnresJonefouroneone$ & $\AmplRatioResJonefouroneone$\\
\TBstrut J2204 & $\freqShortJtwotwozerofour$ & 
$\AmplUnrestULJtwotwozerofour^{+\AmplUnrestULerrpJtwotwozerofour}_{-\AmplUnrestULerrmJtwotwozerofour}\times{10^{\AmplUnrestULexpJtwotwozerofour}}$ & 
-  & 
$\AmplSpndwnJtwotwozerofour$ 	& 
$\epsUnrestULJtwotwozerofour$ 	& - & 
$\AmplRatioUnresJtwotwozerofour$ & - \\
\cutinhead{O3}
\TBstrut J0154 & $\freqShortJzeroonefivefour$	& 
$\AmplUnrestOthreeULJzeroonefivefour^{+\AmplUnrestOthreeULerrpJzeroonefivefour}_{-\AmplUnrestOthreeULerrmJzeroonefivefour}\times{10^{\AmplUnrestOthreeULexpJzeroonefivefour}}$ & 
- & 
$\AmplSpndwnJzeroonefivefour$ 	& 
$\epsUnrestOthreeULJzeroonefivefour$ 	& - & 
$\AmplOthreeRatioUnresJzeroonefivefour$ & -  \\
\TBstrut J0509 & $\freqShortJzerofivezeronine$ & 
$\AmplUnrestOthreeULJzerofivezeronine^{+\AmplUnrestOthreeULerrpJzerofivezeronine}_{-\AmplUnrestOthreeULerrmJzerofivezeronine}\times{10^{\AmplUnrestOthreeULexpJzerofivezeronine}}$ & 
-  & 
$\AmplSpndwnJzerofivezeronine$ 	& 
$\epsUnrestOthreeULJzerofivezeronine$ 	& - & 
$\AmplOthreeRatioUnresJzerofivezeronine$ & -\\
\TBstrut J0709& $\freqShortJzerosevenzeronine$ & 
$\AmplUnrestOthreeULJzerosevenzeronine^{+\AmplUnrestOthreeULerrpJzerosevenzeronine}_{-\AmplUnrestOthreeULerrmJzerosevenzeronine}\times{10^{\AmplUnrestOthreeULexpJzerosevenzeronine}}$ & 
$\AmplRestOthreeULJzerosevenzeronine^{+\AmplRestOthreeULerrpJzerosevenzeronine}_{-\AmplRestOthreeULerrmJzerosevenzeronine}\times{10^{\AmplRestOthreeULexpJzerosevenzeronine}}$  & 
$\AmplSpndwnJzerosevenzeronine$ 	& 
$\epsUnrestOthreeULJzerosevenzeronine$ 	& $\epsRestOthreeULJzerosevenzeronine$ & 
$\AmplOthreeRatioUnresJzerosevenzeronine$ & $\AmplOthreeRatioResJzerosevenzeronine$\\
\TBstrut J0732 & $\freqShortJzeroseventhreetwo$ & 
$\AmplUnrestOthreeULJzeroseventhreetwo^{+\AmplUnrestOthreeULerrpJzeroseventhreetwo}_{-\AmplUnrestOthreeULerrmJzeroseventhreetwo}\times{10^{\AmplUnrestOthreeULexpJzeroseventhreetwo}}$ & 
$\AmplRestOthreeULJzeroseventhreetwo^{+\AmplRestOthreeULerrpJzeroseventhreetwo}_{-\AmplRestOthreeULerrmJzeroseventhreetwo}\times{10^{\AmplRestOthreeULexpJzeroseventhreetwo}}$  & 
$\AmplSpndwnJzeroseventhreetwo$ 	& 
$\epsUnrestOthreeULJzeroseventhreetwo$ 	& $\epsRestOthreeULJzeroseventhreetwo$ & 
$\AmplOthreeRatioUnresJzeroseventhreetwo$ & $\AmplOthreeRatioResJzeroseventhreetwo$\\
\TBstrut J0824 & $\freqShortJzeroeighttwofour$ & 
$\AmplUnrestOthreeULJzeroeighttwofour^{+\AmplUnrestOthreeULerrpJzeroeighttwofour}_{-\AmplUnrestOthreeULerrmJzeroeighttwofour}\times{10^{\AmplUnrestOthreeULexpJzeroeighttwofour}}$ & 
$\AmplRestOthreeULJzeroeighttwofour^{+\AmplRestOthreeULerrpJzeroeighttwofour}_{-\AmplRestOthreeULerrmJzeroeighttwofour}\times{10^{\AmplRestOthreeULexpJzeroeighttwofour}}$  & 
$\AmplSpndwnJzeroeighttwofour$ 	& 
$\epsUnrestOthreeULJzeroeighttwofour$ 	& $\epsRestOthreeULJzeroeighttwofour$ & 
$\AmplOthreeRatioUnresJzeroeighttwofour$ & $\AmplOthreeRatioResJzeroeighttwofour$\\
\TBstrut J1411 & $\freqShortJonefouroneone$ & 
$\AmplUnrestOthreeULJonefouroneone^{+\AmplUnrestOthreeULerrpJonefouroneone}_{-\AmplUnrestOthreeULerrmJonefouroneone}\times{10^{\AmplUnrestOthreeULexpJonefouroneone}}$ & 
$\AmplRestOthreeULJonefouroneone^{+\AmplRestOthreeULerrpJonefouroneone}_{-\AmplRestOthreeULerrmJonefouroneone}\times{10^{\AmplRestOthreeULexpJonefouroneone}}$  & 
$\AmplSpndwnJonefouroneone$ 	& 
$\epsUnrestOthreeULJonefouroneone$ 	& $\epsRestOthreeULJonefouroneone$ & 
$\AmplOthreeRatioUnresJonefouroneone$ & $\AmplOthreeRatioResJonefouroneone$\\
\TBstrut J2204 & $\freqShortJtwotwozerofour$ & 
$\AmplUnrestOthreeULJtwotwozerofour^{+\AmplUnrestOthreeULerrpJtwotwozerofour}_{-\AmplUnrestOthreeULerrmJtwotwozerofour}\times{10^{\AmplUnrestOthreeULexpJtwotwozerofour}}$ & 
-  & 
$\AmplSpndwnJtwotwozerofour$ 	& 
$\epsUnrestOthreeULJtwotwozerofour$ 	& - & 
$\AmplOthreeRatioUnresJtwotwozerofour$ & - \\
%
\cutinhead{O1 O2 O3}
\TBstrut J0154 & $\freqShortJzeroonefivefour$	& 
$\AmplUnrestOoneOtwoOthreeULJzeroonefivefour^{+\AmplUnrestOoneOtwoOthreeULerrpJzeroonefivefour}_{-\AmplUnrestOoneOtwoOthreeULerrmJzeroonefivefour}\times{10^{\AmplUnrestOoneOtwoOthreeULexpJzeroonefivefour}}$ & 
- & 
$\AmplSpndwnJzeroonefivefour$ 	& 
$\epsUnrestOoneOtwoOthreeULJzeroonefivefour$ 	& - & 
$\AmplOoneOtwoOthreeRatioUnresJzeroonefivefour$ & -  \\
\TBstrut J0509 & $\freqShortJzerofivezeronine$ & 
$\AmplUnrestOoneOtwoOthreeULJzerofivezeronine^{+\AmplUnrestOoneOtwoOthreeULerrpJzerofivezeronine}_{-\AmplUnrestOoneOtwoOthreeULerrmJzerofivezeronine}\times{10^{\AmplUnrestOoneOtwoOthreeULexpJzerofivezeronine}}$ & 
-  & 
$\AmplSpndwnJzerofivezeronine$ 	& 
$\epsUnrestOoneOtwoOthreeULJzerofivezeronine$ 	& - & 
$\AmplOoneOtwoOthreeRatioUnresJzerofivezeronine$ & -\\
\TBstrut J0709& $\freqShortJzerosevenzeronine$ & 
$\AmplUnrestOoneOtwoOthreeULJzerosevenzeronine^{+\AmplUnrestOoneOtwoOthreeULerrpJzerosevenzeronine}_{-\AmplUnrestOoneOtwoOthreeULerrmJzerosevenzeronine}\times{10^{\AmplUnrestOoneOtwoOthreeULexpJzerosevenzeronine}}$ & 
$\AmplRestOoneOtwoOthreeULJzerosevenzeronine^{+\AmplRestOoneOtwoOthreeULerrpJzerosevenzeronine}_{-\AmplRestOoneOtwoOthreeULerrmJzerosevenzeronine}\times{10^{\AmplRestOoneOtwoOthreeULexpJzerosevenzeronine}}$  & 
$\AmplSpndwnJzerosevenzeronine$ 	& 
$\epsUnrestOoneOtwoOthreeULJzerosevenzeronine$ 	& $\epsRestOoneOtwoOthreeULJzerosevenzeronine$ & 
$\AmplOoneOtwoOthreeRatioUnresJzerosevenzeronine$ & $\AmplOoneOtwoOthreeRatioResJzerosevenzeronine$\\
\TBstrut J0732 & $\freqShortJzeroseventhreetwo$ & 
$\AmplUnrestOoneOtwoOthreeULJzeroseventhreetwo^{+\AmplUnrestOoneOtwoOthreeULerrpJzeroseventhreetwo}_{-\AmplUnrestOoneOtwoOthreeULerrmJzeroseventhreetwo}\times{10^{\AmplUnrestOoneOtwoOthreeULexpJzeroseventhreetwo}}$ & 
$\AmplRestOoneOtwoOthreeULJzeroseventhreetwo^{+\AmplRestOoneOtwoOthreeULerrpJzeroseventhreetwo}_{-\AmplRestOoneOtwoOthreeULerrmJzeroseventhreetwo}\times{10^{\AmplRestOoneOtwoOthreeULexpJzeroseventhreetwo}}$  & 
$\AmplSpndwnJzeroseventhreetwo$ 	& 
$\epsUnrestOoneOtwoOthreeULJzeroseventhreetwo$ 	& $\epsRestOoneOtwoOthreeULJzeroseventhreetwo$ & 
$\AmplOoneOtwoOthreeRatioUnresJzeroseventhreetwo$ & $\AmplOoneOtwoOthreeRatioResJzeroseventhreetwo$\\
\TBstrut J0824 & $\freqShortJzeroeighttwofour$ & 
$\AmplUnrestOoneOtwoOthreeULJzeroeighttwofour^{+\AmplUnrestOoneOtwoOthreeULerrpJzeroeighttwofour}_{-\AmplUnrestOoneOtwoOthreeULerrmJzeroeighttwofour}\times{10^{\AmplUnrestOoneOtwoOthreeULexpJzeroeighttwofour}}$ & 
$\AmplRestOoneOtwoOthreeULJzeroeighttwofour^{+\AmplRestOoneOtwoOthreeULerrpJzeroeighttwofour}_{-\AmplRestOoneOtwoOthreeULerrmJzeroeighttwofour}\times{10^{\AmplRestOoneOtwoOthreeULexpJzeroeighttwofour}}$  & 
$\AmplSpndwnJzeroeighttwofour$ 	& 
$\epsUnrestOoneOtwoOthreeULJzeroeighttwofour$ 	& $\epsRestOoneOtwoOthreeULJzeroeighttwofour$ & 
$\AmplOoneOtwoOthreeRatioUnresJzeroeighttwofour$ & $\AmplOoneOtwoOthreeRatioResJzeroeighttwofour$\\
\TBstrut J1411 & $\freqShortJonefouroneone$ & 
$\AmplUnrestOoneOtwoOthreeULJonefouroneone^{+\AmplUnrestOoneOtwoOthreeULerrpJonefouroneone}_{-\AmplUnrestOoneOtwoOthreeULerrmJonefouroneone}\times{10^{\AmplUnrestOoneOtwoOthreeULexpJonefouroneone}}$ & 
$\AmplRestOoneOtwoOthreeULJonefouroneone^{+\AmplRestOoneOtwoOthreeULerrpJonefouroneone}_{-\AmplRestOoneOtwoOthreeULerrmJonefouroneone}\times{10^{\AmplRestOoneOtwoOthreeULexpJonefouroneone}}$  & 
$\AmplSpndwnJonefouroneone$ 	& 
$\epsUnrestOoneOtwoOthreeULJonefouroneone$ 	& $\epsRestOoneOtwoOthreeULJonefouroneone$ & 
$\AmplOoneOtwoOthreeRatioUnresJonefouroneone$ & $\AmplOoneOtwoOthreeRatioResJonefouroneone$\\
\TBstrut J2204 & $\freqShortJtwotwozerofour$ & 
$\AmplUnrestOoneOtwoOthreeULJtwotwozerofour^{+\AmplUnrestOoneOtwoOthreeULerrpJtwotwozerofour}_{-\AmplUnrestOoneOtwoOthreeULerrmJtwotwozerofour}\times{10^{\AmplUnrestOoneOtwoOthreeULexpJtwotwozerofour}}$ & 
-  & 
$\AmplSpndwnJtwotwozerofour$ 	& 
$\epsUnrestOoneOtwoOthreeULJtwotwozerofour$ 	& - & 
$\AmplOoneOtwoOthreeRatioUnresJtwotwozerofour$ & -
\enddata
\end{deluxetable*}
\begin{deluxetable}{lcccc}
\tablecaption{O1-O2-O3 targeted searches Bayesian upper limits (unconstrained $\cos
\iota$ priors).
\label{tab:BayesianULs}}
\tablehead{
\colhead{\bf{O1 O2 O3}} & 
\colhead{$h_0^{95\%}$}  &  
\colhead{$\epsilon^{95\%}$} &  
\colhead{${h_0^{95}}/ {h_0^{\textrm{spdwn}}}$} \\ [-0.2cm]
\colhead{\bf{Bayesian}} & 
\colhead{} &  
\colhead{} & 
\colhead{}   \\[-0.1cm]
\colhead{Pulsar} & 
\colhead{} &  
\colhead{} & 
\colhead{}  
}
\startdata
\TBstrut J0154 & 
$\AmplUnrestOonetwothreeBayesULJzeroonefivefour\times{10^{\AmplUnrestOonetwothreeBayesULexpJzeroonefivefour}}$ & 
$\epsUnrestOonetwothreeBayesULJzeroonefivefour$ 	& 
$\AmplOonetwothreeBayesRatioUnresJzeroonefivefour$ & -  \\
\TBstrut J0509  & 
$\AmplUnrestOonetwothreeBayesULJzerofivezeronine\times{10^{\AmplUnrestOonetwothreeBayesULexpJzerofivezeronine}}$ &  
$\epsUnrestOonetwothreeBayesULJzerofivezeronine$ 	& 
$\AmplOonetwothreeBayesRatioUnresJzerofivezeronine$ & -\\
\TBstrut J0709&  
$\AmplUnrestOonetwothreeBayesULJzerosevenzeronine\times{10^{\AmplUnrestOonetwothreeBayesULexpJzerosevenzeronine}}$ & 
$\epsUnrestOonetwothreeBayesULJzerosevenzeronine$ 	& 
$\AmplOonetwothreeBayesRatioUnresJzerosevenzeronine$ \\
\TBstrut J0732 & 
$\AmplUnrestOonetwothreeBayesULJzeroseventhreetwo\times{10^{\AmplUnrestOonetwothreeBayesULexpJzeroseventhreetwo}}$ &  
$\epsUnrestOonetwothreeBayesULJzeroseventhreetwo$ 	&
$\AmplOonetwothreeBayesRatioUnresJzeroseventhreetwo$ \\
\TBstrut J0824 &  
$\AmplUnrestOonetwothreeBayesULJzeroeighttwofour\times{10^{\AmplUnrestOonetwothreeBayesULexpJzeroeighttwofour}}$ & 
$\epsUnrestOonetwothreeBayesULJzeroeighttwofour$ 	& 
$\AmplOonetwothreeBayesRatioUnresJzeroeighttwofour$ \\
\TBstrut J1411 & 
$\AmplUnrestOonetwothreeBayesULJonefouroneone
\times{10^{\AmplUnrestOonetwothreeBayesULexpJonefouroneone}}$ & 
$\epsUnrestOonetwothreeBayesULJonefouroneone$ 	& 
$\AmplOonetwothreeBayesRatioUnresJonefouroneone$ \\
\TBstrut J2204 & 
$\AmplUnrestOonetwothreeBayesULJtwotwozerofour
\times{10^{\AmplUnrestOonetwothreeBayesULexpJtwotwozerofour}}$ &  
$\epsUnrestOonetwothreeBayesULJtwotwozerofour$ 	& 
$\AmplOonetwothreeBayesRatioUnresJtwotwozerofour$ \\
\enddata
\end{deluxetable}

If we assume that the neutron star is a triaxial ellipsoid spinning around a principal moment of inertia axis ${I_{zz}}$, and that the continuous wave emission is due to an ellipticity
\begin{equation}
\varepsilon= {{I_{xx}-I_{yy}}\over{ I_{zz}}},
\label{eq:epsilonDef}
\end{equation}
based on the intrinsic gravitational wave amplitude upper limits $h_0^{95\%}$, we can exclude neutron star deformations above a $\varepsilon^{95\%}$ level.  The ellipticity needed for a neutron star at a distance $D$, spinning at $f/2$, to produce continuous gravitational waves with an intrinsic amplitude on Earth of $h_0$ is \citep{Jaranowski:1998qm, Gao:2020zcd}:
\begin{equation}
\begin{split}
\varepsilon &=2.4 \times 10^{-7} ~\left( {h_0\over{1\times10^{-26}}}\right ) \times \\
&\left ( {D\over{1~\textrm{kpc}}}\right ) \left ({{\textrm{200~Hz}}\over f} \right )^2 \left ({10^{38}~{\textrm{kg m}}^2\over I_{zz}} \right ).\\
\end{split}
\label{eq:epsilon}
\end{equation}
The ellipticity $\varepsilon^{95\%}$ upper limits are given in Table \ref{tab:ULs}. 


\subsection{Discussion}
\label{sec:discussion}

We have searched for continuous gravitational waves from seven pulsars that have not been targeted before. We use all the publicly available Advanced LIGO data, namely from the O1, O2 and O3 science runs. 

We find no evidence of a gravitational wave signal at a detectable level. The posterior probability distribution for PSR J0709 is peaked slightly off-zero, but this could well be a noise fluctuation as well as due to spectral contamination. Especially at the lower frequencies it is not uncommon to find these posteriors, see for example Figure 3 of \citep{Abbott:2020lqk} showing the results for the Vela Pulsar from the search at $\approx$ 22 Hz. The $\F$-statistic results for the coherent O1O2O3 search are insignificant, which indicates that a coherent signal during all the observation is not detected. On the other hand PSR J0709 is one of the only two pulsars for which the EM timing overlaps with all the LIGO runs, so we are most confident of the used template waveform.

For more than half of the pulsar sample, our searches probe ellipticities  $\lesssim  3 \times 10^{-7}$, which could be sustained by neutron star crusts \citep{JohnsonMcDaniel:2012wg,Gittins:2020cvx,Bhattacharyya:2020paf}. Our tightest ellipticity bound amounts to $1.7\times 10^{-8}$ ($1.5\times 10^{-8}$ from the Bayesian analysis), for PSR J0154. The remaining four pulsars are more distant and/or spin slower, which yields less constraining ellipticity upper limits.  For the pulsar PSR J0824, assuming a canonical moment of inertia of $10^{38} {\textrm{kg m}}^2$, our upper limits are within a factor of 3.8 (5.8) of the spindown upper limit, for an unconstrained and constrained $\cos\iota$ prior respectively. The actual moment of inertia of the star may differ from the canonical one up by a factor of a few. These are physically interesting ellipticity ranges \citep{Woan:2018tey}, and showcase the potential for this type of search.


\acknowledgments
All the computational work for these searches was performed on the ATLAS cluster at AEI Hannover. We thank Carsten Aulbert and Henning Fehrmann for their support. \\
We would like to especially thank the instrument-scientist and engineers of LIGO whose amazing work has produced detectors capable of probing gravitational waves so incredibly small.\\
This research has made use of data, software and/or web tools obtained from the Gravitational Wave Open Science Center (https://www.gw-openscience.org/ ), a service of LIGO Laboratory, the LIGO Scientific Collaboration and the Virgo Collaboration. LIGO Laboratory and Advanced LIGO are funded by the United States National Science Foundation (NSF) as well as the Science and Technology Facilities Council (STFC) of the United Kingdom, the Max-Planck-Society (MPS), and the State of Niedersachsen/Germany for support of the construction of Advanced LIGO and construction and operation of the GEO600 detector. Additional support for Advanced LIGO was provided by the Australian Research Council. Virgo is funded, through the European Gravitational Observatory (EGO), by the French Centre National de Recherche Scientifique (CNRS), the Italian Istituto Nazionale di Fisica Nucleare (INFN) and the Dutch Nikhef, with contributions by institutions from Belgium, Germany, Greece, Hungary, Ireland, Japan, Monaco, Poland, Portugal, Spain.

\bibliography{paperBibApJ}{}
\bibliographystyle{aasjournal}
\bibstyle{aasjournal}

\appendix

\section{Additional plots}
\begin{figure}[h!]
\includegraphics[width=1\textwidth]{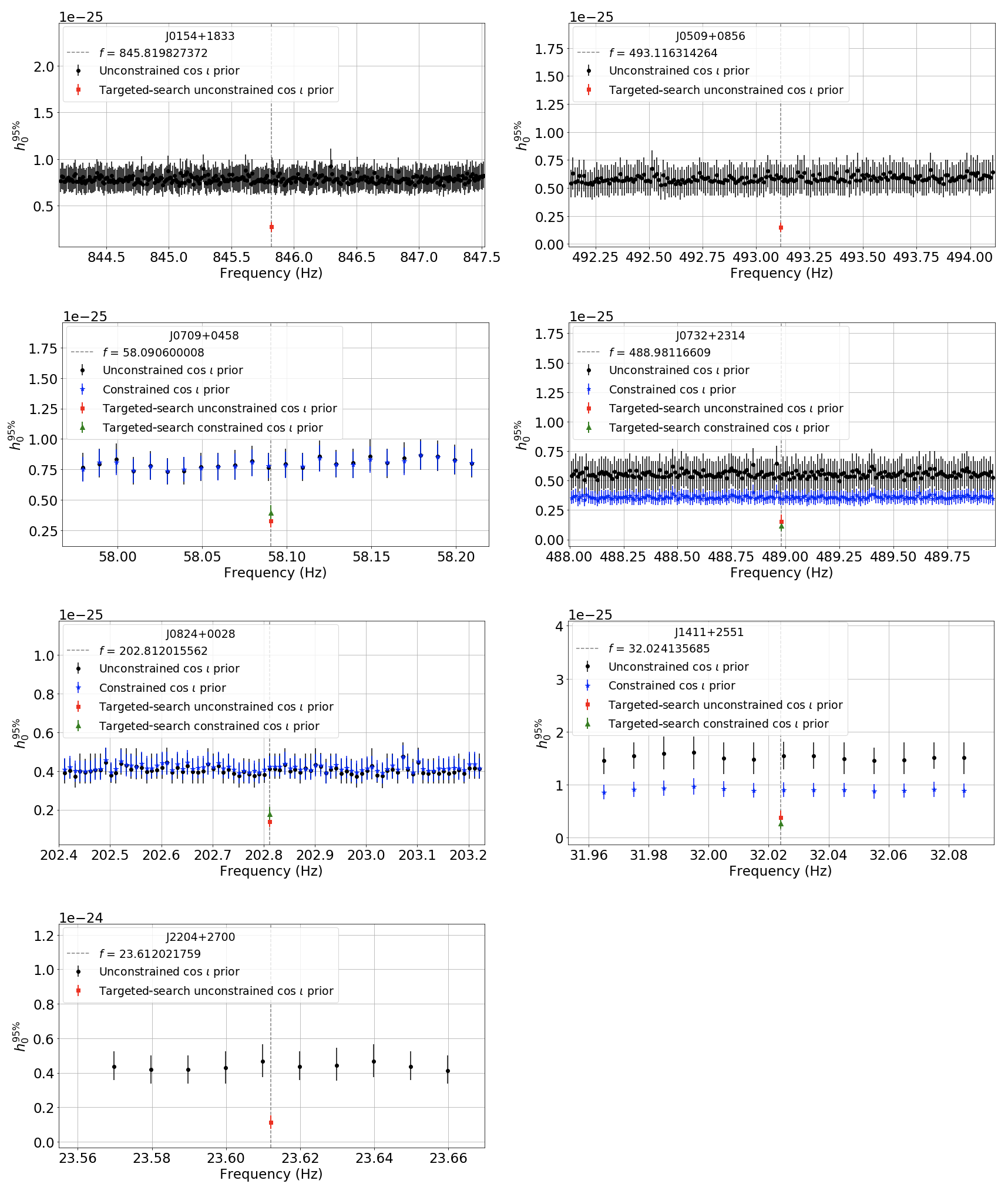}
\caption{O1O2 data upper limits on the gravitational wave amplitude in each 10 mHz frequency sub-band searched, based on the most significant result in that 10 mHz sub-band. We also show the targeted search results, which are the lower points at the central frequency, which is twice the pulsar rotation frequency. 
}
\label{fig:h0ULs}
\end{figure}

\section{Gravitational waveform parameters}
\hspace{-5cm}
\begin{deluxetable}{lcccccccccc}
\rotate
\centerwidetable
\tablecaption{Gravitational waveform parameters
\label{tab:pulsarParams}
}
\tablehead{
\colhead{PULSAR} & \colhead{GW} & \colhead{GW} & \colhead{RA} & \colhead{DEC} &  \colhead{Epoch } & \colhead{Proj. semi-major} & \colhead{Binary} & \colhead{Eccentricity} & \colhead{Arg. of } & \colhead{Distance} \\
& \colhead{\small{frequency $f$}} & \colhead{ \small{freq. derivative $\fdot$}} & \colhead{} & \colhead{} & ${\tau_0}_{\mathrm{SSB}}$ &\colhead{axis, a $\sin \textrm{i}$} &\colhead{ Period} & & \colhead{Periastron $\omega$} & \colhead{} \\
& \colhead{[Hz]} & \colhead{[Hz/s]} & \colhead{[rad]} & \colhead{[rad]} &  \colhead{ [MJD]} & \colhead{[light-s]} & \colhead{[s]} & & \colhead{[rad]} & \colhead{[pc]} \\
}
\startdata
\TBstrut J2204 & $\freqShortJtwotwozerofour$ 	& $\fdotJtwotwozerofour$ 	& 
$\raJtwotwozerofour$ 	& $\decJtwotwozerofour$ 	&   
$\pepochJtwotwozerofour$ & $\asiniJtwotwozerofour$ & $\pbJtwotwozerofour$	& $\eccJtwotwozerofour$	& $\omegaJtwotwozerofour$	& 
$\DistJtwotwozerofour$\\
\TBstrut J1411 & $\freqShortJonefouroneone$ 	& $\fdotJonefouroneone$ 	& 
$\raJonefouroneone$ 	& $\decJonefouroneone$ 	&   
$\pepochJonefouroneone$ & $\asiniJonefouroneone$ & $\pbJonefouroneone$	& $\eccJonefouroneone$	& $\omegaJonefouroneone$&  $\DistJonefouroneone$   \\
%
\TBstrut J0709 & $\freqShortJzerosevenzeronine$ 	& $\fdotJzerosevenzeronine$ 	& 
$\raJzerosevenzeronine$ 	& $\decJzerosevenzeronine$ &
$\pepochJzerosevenzeronine$ & $\asiniJzerosevenzeronine$ & $\pbJzerosevenzeronine$ & $\eccJzerosevenzeronine$	& $\omegaJzerosevenzeronine$& 
$\DistJzerosevenzeronine$    \\
\TBstrut J0824 & $\freqShortJzeroeighttwofour$ 	& $\fdotJzeroeighttwofour$ 	& 
$\raJzeroeighttwofour$ 	& $\decJzeroeighttwofour$ 	&   
$\pepochJzeroeighttwofour$ & $\asiniJzeroeighttwofour$ & $\pbJzeroeighttwofour$	& $\eccJzeroeighttwofour$	& $\omegaJzeroeighttwofour$	& 
$\DistJzeroeighttwofour$    \\
\TBstrut J0732 & $\freqShortJzeroseventhreetwo$ 	& $\fdotJzeroseventhreetwo$ 	& 
$\raJzeroseventhreetwo$ 	& $\decJzeroseventhreetwo$ 	&   
$\pepochJzeroseventhreetwo$ &  $\asiniJzeroseventhreetwo$ & $\pbJzeroseventhreetwo$ & $\eccJzeroseventhreetwo$ & $\omegaJzeroseventhreetwo$ & 
$\DistJzeroseventhreetwo$    \\
\TBstrut J0509 & $\freqShortJzerofivezeronine$ & $\fdotJzerofivezeronine$  &
$\raJzerofivezeronine$ 	& $\decJzerofivezeronine$ 	&   
$\pepochJzerofivezeronine$ & $\asiniJzerofivezeronine$ & $\pbJzerofivezeronine$ & $\eccJzerofivezeronine$ &$\omegaJzerofivezeronine$	& 
$\DistJzerofivezeronine$    \\
\TBstrut J0154 & $\freqShortJzeroonefivefour$	& $\fdotJzeroonefivefour$ 	& 
$\raJzeroonefivefour$ 	& $\decJzeroonefivefour$ 	&   
$\pepochJzeroonefivefour$ & - & -	& -	& -	& 
$\DistJzeroonefivefour$   \\
\enddata
\end{deluxetable}

\end{document}